%% file: main.tex
\documentclass{article} %
\usepackage{iclr2025_conference,times}

\input{sections/preamble}

\newcommand{\titleString}{\textcolor{myblue}{VideoPanda}: \textcolor{myblue}{Video Pan}oramic \textcolor{myblue}{D}iffusion with Multi-view \textcolor{myblue}{A}ttention %
}

\title{\titleString}

\def\mystrut{\rule{0pt}{1.0\normalbaselineskip}}
\author{
\begin{tabular}{@{}l}
Kevin Xie\thanks{First and second author contributed equally. Correspondence to: \texttt{\scriptsize chxie@nvidia.com}}\quad Amirmojtaba Sabour$^*$\quad Jiahui Huang\quad Despoina Paschalidou\mystrut \\
Greg Klar\quad Umar Iqbal\quad Sanja Fidler\quad Xiaohui Zeng\mystrut \\
\end{tabular}\\
NVIDIA\mystrut
}

\newcommand{\ourmodel}{VideoPanda\xspace}

\iclrfinalcopy %
\begin{document}

\maketitle

\begin{figure}[htbp]
    \makebox[\linewidth][c]{%
        \includegraphics[width=1.02\linewidth,trim={0.0cm 0 0 0},clip]{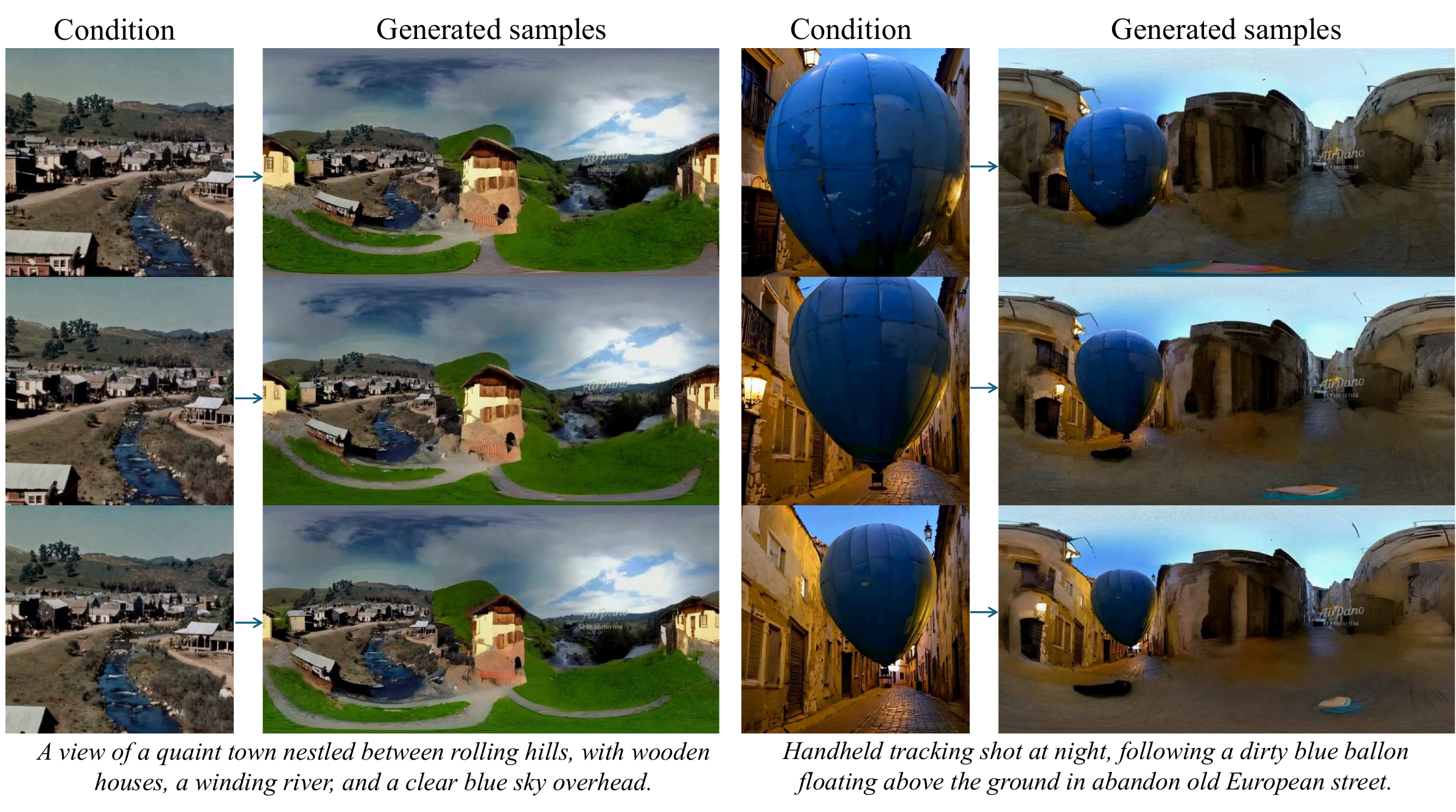}%
    }
    \captionof{figure}[Short caption]{\small Generated samples conditioned on a single-view video and text prompt. Both single-view video inputs were generated using existing video generation models~\citep{sora,runway}.
    Auto-regressive generation is applied to extend the video length. %
    }
    \label{fig:teaser-top}
\end{figure}

\input{sections/abstract}

\input{sections/introduction}

\input{sections/related_work}

\input{sections/method}

\input{sections/experiment}

\input{sections/conclusion}

\bibliography{main.bib}
\bibliographystyle{iclr2025_conference}

\appendix
\input{sections/appendix}

\end{document}

%% file: sections/preamble.tex
\usepackage{makecell}
\usepackage{booktabs}
\usepackage{multirow}

\usepackage{siunitx}
\usepackage{xcolor}         %
\usepackage{wrapfig}
\usepackage{subcaption}
\usepackage{mwe}
\usepackage{graphbox} %
\usepackage{pgfplots} 
\usepackage{tabularx} %
\usepackage{amsmath}
\usepackage{hyperref}
\usepackage{enumitem}
\usepackage{url}
\usepackage{xspace}
\usepackage{floatflt}
\usepackage{titletoc} %

\hypersetup{
    colorlinks=true,    %
    linkcolor=blue,     %
    citecolor=blue,     %
    urlcolor=blue,      %
    pdfborder={0 0 0},  %
}

\input{math_commands.tex}

\newif\ifcomments
\commentstrue

\definecolor{nvgreen}{HTML}{76B900}
\definecolor{myblue}{HTML}{5364cc}

\usepackage[capitalize,nameinlink]{cleveref}

\crefname{section}{\S}{\S\S}
\crefname{subsection}{\S}{\S\S}
\crefname{conj}{Conj.}{Conj.}

\crefname{algorithm}{\text{Alg.}}{\text{Alg.}}
\crefname{assumption}{\textbf{H}}{\textbf{H}}
\crefname{equation}{\text{Eq}}{\text{Eq}}
\crefname{definition}{\text{Dfn.}}{\text{Dfn.}}
\crefname{lemma}{\text{Lemma}}{\text{Lemma}}
\crefname{dfn}{\text{Dfn.}}{\text{Dfn.}}
\crefname{thm}{\text{Thm.}}{\text{Thm.}}
\crefname{fig}{\text{Fig.}}{\text{Fig.}}
\crefname{table}{\text{Tab.}}{\text{Tab.}}

%% file: math_commands.tex
\usepackage{amsmath,amsfonts,bm}

\def\eqref#1{equation~\ref{#1}}

\def\1{\bm{1}}

\def\eps{{\epsilon}}

\DeclareMathAlphabet{\mathsfit}{\encodingdefault}{\sfdefault}{m}{sl}
\SetMathAlphabet{\mathsfit}{bold}{\encodingdefault}{\sfdefault}{bx}{n}

%% file: sections/abstract.tex
\begin{abstract}

High resolution panoramic video content is paramount for immersive experiences in Virtual Reality, but is non-trivial to collect as it requires specialized equipment and intricate camera setups. 
In this work, we introduce \ourmodel, a novel approach for synthesizing $360^\circ$ videos conditioned on text or single-view video data. \ourmodel leverages multi-view attention layers to augment a video diffusion model, enabling it to generate consistent multi-view videos that can be combined into immersive panoramic content. \ourmodel is trained jointly using two conditions: text-only and single-view video, and supports autoregressive generation of long-videos. 
To overcome the computational burden of multi-view video generation, we randomly subsample the duration and camera views used during training and show that the model is able to gracefully generalize  to generating more frames during inference.
Extensive evaluations on both real-world and synthetic video datasets demonstrate that \ourmodel generates more realistic and coherent $360^\circ$ panoramas across all input conditions compared to existing methods.
Visit the project website at \url{https://research.nvidia.com/labs/toronto-ai/VideoPanda/} for results.

\end{abstract}

%% file: sections/introduction.tex
\section{Introduction}
A key aspect of achieving true immersion in a virtual environment is allowing users to look around freely, by rotating their head and exploring their surroundings from all possible angles. To enable such experiences, it is essential to have access to high-quality and high-resolution panoramic videos. 
However, recording such videos is both expensive and time-consuming, as it requires intricate camera setups and specialized equipment. As a result, the available panoramic video content on platforms such as YouTube or Vimeo remains limited compared to single-view videos. In this work, we aim to address this issue, by developing a generative model capable of synthesizing panoramic videos either from text prompts or by expanding single-view videos (either generated from models like Sora~\citep{sora} or recorded) into panoramic format. We consider this an essential step towards making immersive content more accessible and scalable.

Recently, diffusion models have shown remarkable success in generating images~\citep{ho2022imagen,blattmann2023stable}, 3D models~\citep{shi2023MVDream,poole2022dreamfusion}, and videos~\citep{sora,blattmann2023videoldm} from text prompts.  %
Despite their promising capabilities,  generation of  panoramic videos using diffusion models presents significant challenges, mainly due to the scarcity of high-quality panoramic video datasets.
Furthermore, while substantial progress has been made towards advancing standard video generation pipelines~\citep{girdhar2023emuvideo,hong2022cogvideo,chen2024videocrafter2,opensora}, very few works have attempted to apply these techniques to panoramic video generation. Existing methods are either limited to specific domains such as driving scenarios~\citep{wen2024panacea,wu2024drivescape,li2023drivingdiffusion,zhao2024drive,liu2024mvpbev} or restricted to generating static scenes~\citep{wu2023panodiffusion,zhang2024taming}.
360DVD~\citep{wang2024360dvd} directly generates equirectangular panorama video (with text condition), which presents a large domain gap to base model pretrained on perspective view.
We perform an extensive comparison to 360DVD in the text-conditional setting and demonstrate our improved visual quality.

In this paper, we introduce \textbf{\ourmodel}, a novel approach capable of generating high-quality panoramic videos from text prompts and single-view video, as well as creating long video using auto-regression. 
Our approach builds on existing video diffusion models by adding multi-view attention layers to generate consistent multi-view outputs. Doing so ensures that the output domain (perspective images) remains close to the original training distribution of the pretrained video model (as opposed to directly generating equirectangular projections), which helps in maintaining video quality while generating multiple views. The resulting views are then seamlessly stitched together to create a cohesive panoramic video. 
We evaluate our model on a diverse set of data domains, including both real and synthetic videos, and demonstrate its superior performance and quality compared to previous approaches, both quantitatively and qualitatively. Additionally, a user study indicates that the majority of participants prefer our generated videos over those from other baseline models.
In summary, we make the following contributions:
\begin{itemize}[leftmargin=*,noitemsep,topsep=0pt]
    \item %
    We identify the value of panoramic video generation by allowing users to input single-view videos as a condition -- a widely available modality, and present a multi-view video architecture capable of generating plausible panoramic videos. 
    \item We demonstrate that our model can be jointly trained for text-conditioning, video-conditioning, and autoregressive settings by randomizing the conditioning type, leading to improved results and enabling the generation of long panoramic videos. %
    \item When extending the video model to multi-view, the number of generated image frames greatly increases. We overcome the inherent computational burden associated by randomly subselecting the number of views and frames and show that it gracefully generalizes to video with long duration and more views during inference.
    
\end{itemize}

%% file: sections/related_work.tex
\section{Related Work}
\subsection{Image and Video Diffusion Models}
Diffusion models~\citep{ho2020denoising} have demonstrated remarkable success in generating high-quality images~\citep{Karras2022edm,Karras2024edm2,pernias2023wurstchen,Hoogeboom2023simplediffusion,ho2022imagen} and videos~\citep{girdhar2023emuvideo,hong2022cogvideo,blattmann2023stable,blattmann2023videoldm,sora,guo2023animatediff,chen2023gentron,gupta2023photorealistic} from text prompts. To reduce the computational cost of generating high-dimensional data such as images and videos, latent diffusion models ~\citep{rombach2022high} (LDMs) proposed to first encode the data into a compressed latent space using a variational autoencoder (VAE)~\citep{kingma2013auto}, and then conduct the diffusion in this lower dimensional space. 
These models have been proven highly effective for a wide range of downstream tasks such as inpainting~\citep{lugmayr2022repaint}, controllable generation~\citep{zhang2023controlnet}, customized generation~\citep{ruiz2023dreambooth}, and image/video editing~\citep{kawar2023imagic,molad2023dreamix} etc.

\subsection{Multi-View Image Generation}
Building on the success of diffusion models for 2D image generation, they have been increasingly adapted also for multi-view image generation. However, due to the limited availability of real-world multi-view training data, several recent approaches~\citep{shi2023MVDream,long2024wonder3d,liu2023syncdreamer,liu2023zero} attempted to  fine-tune pretrained image generation models like Stable Diffusion \citep{rombach2022high} to support multi-view generation. 
Such approaches can be roughly categorized into two categories: object-centric and scene-centric approaches. 

Object-centric models focus primarily on generating images of objects where all cameras are inward-facing, looking at a single object from different viewing directions. Examples of such approaches include \citep{kant2024spad,kong2024eschernet,shi2023zero123++, shi2023MVDream,tang2024mvdiffusion++,voleti2024sv3d, wang2023imagedream}. More recently, several object-centric generative models explored incorporating custom attention mechanisms~\citep{hu2024mvdfusion,huang2023epidiff,kant2024spad,li2024era3d} to aggregate view-specific information across multiple views.
Notable among these, CAT3D~\citep{gao2024cat3d} trains a model that generates novel views of an inward-focused scene from one or more input views, allowing for 3D reconstruction from a single image. However, these methods often focus on single-object scenes, which limits their applicability to more complex environments.

The second line of work seeks to generate realistic multi-view images of entire scenes, using outward-facing cameras to capture different viewing directions  and produce panoramas. For instance, PanoDiffusion~\citep{wu2023panodiffusion} is trained on equirectangular projections of $360^\circ$ panoramic images, and relies on inpainting during inference to extend the input images into complete panoramas. 
Building on this, PanFusion~\citep{zhang2024taming} adds an additional branch to Stable Diffusion, enabling the simultaneous generation of panoramas and multi-view images. 
MVDiffusion~\citep{Tang2023mvdiffusion} introduces correspondence-aware attention (CAA) layers, where each point attends only to other points within its local neighborhood. %
More recently \citep{yuan2024camfreediff,wang2023panodiff} proposed predicting the homography between input views and use a diffusion model to generate the unseen regions of the panorama. 
Lastly, LayerPano3D~\citep{yang2024layerpano3d} combines multi-view and inpainting models to generate multi-layer panoramas, allowing for somewhat limited exploration within the scene boundaries. Other notable works in this area include ~\citep{li2024genrc,zhou2024holodreamer,hara2024magritte,liu2024panofree}.

\subsection{Multi-view and Panorama Video Generation}
The emergence of powerful open-source video diffusion models~\citep{blattmann2023stable,chen2024videocrafter2,opensora} gave rise to the development of several approaches aimed at augmenting them with multi-view capabilities~\citep{watson2024controlling} and extending them to generate $360^\circ$ panoramic videos. 
For example, 360DVD~\citep{wang2024360dvd} builds upon a pretrained text-to-video model~\citep{guo2023animatediff} by adding a 360-adapter and fine-tuning it on equirectangular projections of panoramic videos. This enables the creation of $360^\circ$ videos from a text inputs, with the option to condition on optical flow videos.
Generative Camera Dolly~\citep{van2024generative} extends the image-conditional Stable Video Diffusion~\citep{blattmann2023stable} into a video-to-video model. Given an input video of a scene, \citep{blattmann2023stable} can generate a synchronized video from a different camera trajectory. 
4K4DGEN~\citep{Li20244K4DGenP4} draws inspiration from MultiDiffusion~\citep{muller2024multidiff} and introduces a training-free method that denoises multiple views of a spherical panorama simultaneously.
Most similar to our method is Panacea~\citep{wen2024panacea}, which is inspired byVideoLDM~\citep{blattmann2023stable} and extends StableDiffusion by adding multi-view and temporal attention layers, trained on multi-view driving videos. 
Notably, Panacea relies on a dynamic birds' eye view (BEV) representation as conditioning, which is most commonly available in the case of driving scenes, thus effectively limiting its applications to driving scenes.

%% file: sections/method.tex
\section{Method}
\begin{figure}
    \centering
    \includegraphics[width=\linewidth]{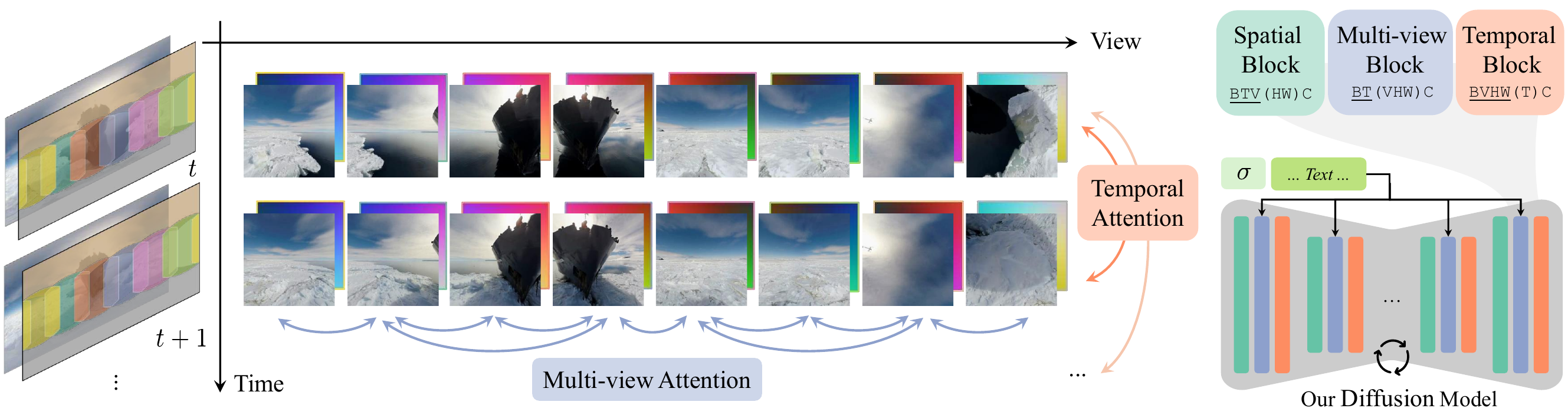}
    \caption{We divide the equi-rectangular video into 8 perspective views via projection. Our diffusion model consists of interleaved spatial, multi-view, and temporal blocks, conditioned on text prompts. Attention is used to propagate information through the multi-view videos to ensure consistency. The input views are embedded using the ray directions as visualized by the color map behind the perspective images.}
    \label{fig:pipeline}
\end{figure}

In this work, we introduce \ourmodel, a multi-view video diffusion model capable of generating long panoramic $360^\circ$ videos from a text prompt or a perspective video. 
Below, we describe our multi-view video diffusion model (\cref{method:model_design}), detail the model training strategy (\cref{method:training_strategy}), and finally describe the approach for auto-regressively generating long videos (\cref{method:autoregressive}). 
\cref{fig:pipeline} provides an overview of our general model design. 

\subsection{Model Design} \label{method:model_design}
We train a multi-view video diffusion model that, given a text prompt and an optional set of conditioning frames, is able to jointly generate multiple multi-view consistent videos of different view directions that together cover a full $360$\textdegree~panoramic video.

Our architecture builds on video latent diffusion models (VLDM) \citep{blattmann2023videoldm} by incorporating multi-view attention layers inspired by MVDream~\citep{shi2023MVDream} and injecting view direction embeddings into the model. Specifically, we add 3D multi-view self-attention layers that perform self-attention across images from different views at each frame of the video. These layers are combined with the existing 2D self-attention layers in a residual manner using zero-initialized convolutions, similar to ControlNets~\citep{zhang2023controlnet}. To provide the model with an understanding of viewing directions, we use ray direction representations that are the same height and width as the latent representations and encode the ray directions at each spatial location, following~\citep{gao2024cat3d}. These rays are defined relative to the camera pose of the first view, and are invariant to global 3D translations and rotations. 
The view embeddings are concatenated channel-wise with their corresponding latents and are fed into the model at the first layer using zero-initialized convolutions.

Given a set of target and optional conditioning frames of size $512 \times 512 \times 3$, each image is encoded into a latent representation of size $64 \times 64 \times 4$ using a variational autoencoder (VAE)~\citep{kingma2013vae}. To enable conditioning on specific frames, we adopt the approach from CAT3D~\citep{gao2024cat3d}. 
During training, the latents corresponding to the non-conditioned views are noised according to the diffusion process, while the latents of the conditioning frames are kept mostly clean. 
Following prior work~\citep{ho2021cascaded}, to improve robustness and prevent overfitting, we use \textbf{noise augmentation} by adding a small amount of noise $\sigma$ to the input conditioning latents and pass this value $\sigma$ to the model as well.
A binary mask is concatenated channel-wise to distinguish between the input conditioning latents and the target frames to be predicted. The diffusion model is then trained to learn the joint distribution of these latent representations conditioned on the inputs. We incorporate classifier-free guidance (CFG)~\citep{ho2022cfg} by randomly dropping the conditioning frames with a probability of 10\% during training.

Finally, similar to prior works~\citep{Hoogeboom2023simplediffusion}, we observe improved performance when shifting the noise schedule towards higher noise levels, as our model generates more image frames than the base video model. Please see \cref{appendix:ablating_shiftsnr} for more details. We also find that using a $v$-prediction objective~\citep{salimans2022progressivedistillation} leads to more stable training compared to $\eps$-prediction, particularly with high-noise schedules.

\subsection{Training Strategy} \label{method:training_strategy}

We initialize the model from the pretrained text-to-video diffusion model VideoLDM which was presented in ``Align Your Latents''~\citep{blattmann2023videoldm}. Following prior works~\citep{shi2023MVDream}, the weights of the multi-view attention layers are initialized with the same weights as the existing 2D self-attention layers to accelerate training.

As we want to adjust the noise schedule (shifting toward higher noise levels) and change the model parametrization from $\epsilon$-prediction to $v$-prediction without overfitting the model to our limited panorama videos, we train our model in two stages. In the first stage, we finetune the single-view text-to-video model from the existing checkpoint, adapting it to the new noise schedule and loss objective. This stage is performed on a subset of the original pretraining data with standard captioned videos of 16 frames and requires minimal training time, as the model adapts quickly to these changes.
In the second stage, we freeze the spatial layers of the video model and finetune the rest using multi-view video data.

During training, we randomize both the number of views and video frames to enhance the model’s generalization and prevent overfitting to the limited $360^\circ$ video data, effectively using this as a form of data augmentation. The model is trained to generate multi-view video sequences represented as view-frame matrices of varying sizes, such as $3 \times 16$, $4 \times 12$, $6 \times 8$, and $8 \times 6$, where the first dimension refers to the number of views and the second to the number of frames. We refer to this randomization as \textbf{random matrix} going forward. This allows the model to generalize to new view-frame combinations, like $8 \times 16$ matrices, during inference—configurations that couldn’t fit in GPU memory during training.

\begin{figure}[t]
\centering
    \includegraphics[width=\textwidth]{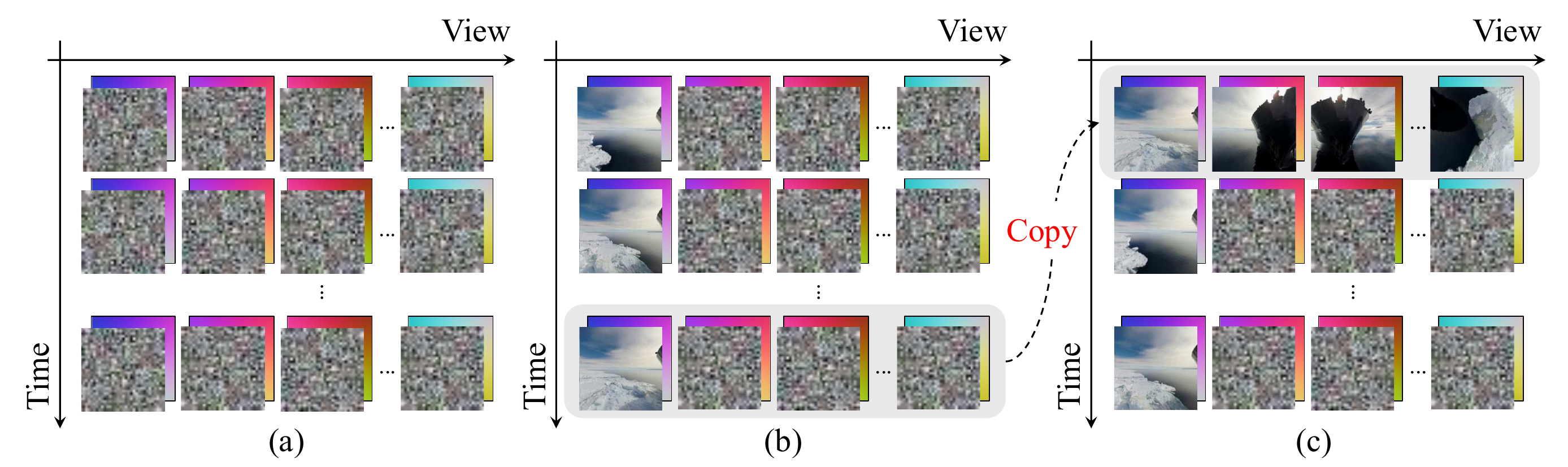}
    \caption{The model is trained using three frame conditioning regimes. (a) No image conditions and the initial inputs are pure noise; (b) Conditioning only on the first view of the video; (c) Conditioning on the first frame and first views for auto-regressive video generation. At inference time, we autoregressively condition on long videos by using conditioning (b) to generate the first window and subsequently using the last multi-view images row from the previous time step (the shaded region) as the first row input to our model using condition-type (c).  %
    }
    \label{fig:input}
\end{figure}

To handle multiple conditioning scenarios, we train a single general model that can generate multi-view videos conditioned on text, video, or a combination of video and the first frame's multi-view images for autoregressive generation using a \textbf{multi-task} training strategy. Specifically, the binary mask is randomized to reflect these different conditioning setups: all zeros (text conditioning), the first column of ones with zeros elsewhere (video conditioning), or both the first row and first column set to ones (autoregressive generation), with equal probability. See \cref{fig:input} for a visualization of the different types of conditioning.

\subsection{Autoregressive Generation of Long Videos}
\label{method:autoregressive}

To generate long panoramic videos, we use an autoregressive approach (see \cref{fig:input}). Initially, conditioned on the first 16-frames of the input video, the model generates an $8 \times 16$ view-frame matrix. For subsequent frames, the model is conditioned on the next 15 new frames of the video (a column) and the last frame from all 8 views (a row) generated in the previous step. This iterative process allows us to generate long, coherent video sequences with smooth transitions and consistent motion.

Autoregressive generation, however, tends to accumulate errors over time, leading to a gradual degradation in image quality and noticeable blurring after a few iterations. The noise augmentation introduced in \cref{method:model_design} helps mitigate this issue, consistent with findings from prior work~\citep{valevski2024gamengen}. This noise augmentation serves two purposes: it acts as a data augmentation technique to improve generalization, and it allows the model to self-correct by learning to recover clean information from noisy samples generated in previous iterations. Please see \cref{appendix:ablating_noise_aug} for details.

%% file: sections/experiment.tex
\input{figures_tex/ours_vs_dvd360}
\input{figures_tex/ours_vs_mvdiffusion}

\section{Experiments}
In this section, we explain the details of our experimental setting and our methodology for evaluations. We then present qualitative and quantitative comparisons to assess our models efficacy against baselines in text and video-conditional generation, demonstrate our models extension to long video generation and ablate key components of our training strategy. Additional training details are included in \cref{appendix:extra_train_detail}.

\subsection{Data}
\paragraph{Training Data.}
We train our model on the WEB360~\citep{wang2024360dvd} dataset, which contains 2,114 panorama video clips with automatically generated captions. Each clip is 100 frames in length, totalling approximately 3 hours of footage that predominantly features panning shots of outdoor scenery.

\paragraph{Evaluation Data.}
For the video conditioning task, we evaluate our method on both in-distribution and out-of-distribution data:
\begin{itemize}[leftmargin=*,noitemsep,topsep=0pt]
    \item In-distribution evaluation data: We gather 100 unseen panorama video clips from Youtube and extract 90 FOV horizontal perspective views for the input conditioning. Prompts are obtained by captioning the middle frame of the conditioning video using CogVLM~\citep{cogvlm}. We use these captions only when evaluating the in-distribution text-conditional generation.
    \item Out-of-distribution text-conditional input: We use prompts from the popular VBench video generation benchmark.
    \item Out-of-distribution video-conditional input: We use generated videos from models including SORA~\citep{sora}, Runway~\citep{runway} and Luma~\citep{luma}. These videos are cropped and resized to a resolution of $512\times 512$ and treated  as horizontal side views. When available, we use the original prompt; otherwise, we caption the middle frame with CogVLM.
\end{itemize}
Since the out-of-distribution condition inputs do not originate from 360 videos, we cannot compute metrics that require ground truth images, such as pairwise FVD \citep{Unterthiner2018ARXIV} and the reconstruction metrics. We include more details about the out of distribution evauation and their quantitative evaluation in \cref{appendix:ood_eval}.

\paragraph{Processing.} 

\begin{wrapfigure}{r}{0.28\textwidth}
    \centering
    \includegraphics[width=\linewidth]{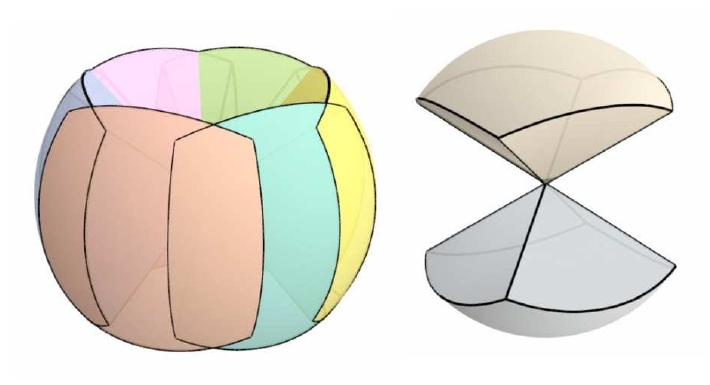}
    \caption{A visualization of the 8 frames used during training, consisting of 6 horizontal views with 90 FOV and 2 views for the top/bottom with 100 FOV}
    \label{fig:emb}
\end{wrapfigure}

We first convert the equirectangular video data into multiple perspective views with overlap. A visualization is shown in \cref{fig:emb}.
Similar to MVDiffusion we cover the horizontal side views with multiple perspective views of 90\textdegree~FOV at 0\textdegree~elevation. We empirically observed that the excessive amount of overlap stemming from the use of 8 horizontal views was unnecessary and thus we only use 6 views instead. These are evenly spaced in azimuth in offsets of 60\textdegree. We also explored using just 4 views which results in no overlaps between views similar to a cubemap representation but found that it was more difficult for the model to maintain consistency between views without overlaps.
Additionally, to obtain a full panorama we add two perspective views looking straight up (90\textdegree~elevation) and down (-90\textdegree~elevation) to cover the `sky' and `ground' views. We increase the FOV for these two to 100\textdegree~which is large enough to cover all pixels in the panorama when combined with the 6 side views.

\subsection{Inference}

Unless otherwise specified, we use a DDIM sampler with 25 steps and
classifier-free guidance (CFG) to improve generation quality. %
In the text-conditional setting, we use a CFG scale of 8.0.
For the video-conditional setting, we use a CFG scale of 4.0, where the unconditional score prediction does not take text nor video as input.

To facilitate fair evaluation, we use a common equirectangular format with resolution $512\times 1024$ for a $16$ frame long panoramic video and compose our multi-view results into it by warping each of the images with bicubic interpolation. The pixel values in regions with overlap between views are uniformly averaged.
When evaluating our generations, we either directly evaluate the stitched equirectangular video or following MVDiffusion, we crop 8 horizontal perspective view videos from it, since some metrics are more naturally evaluated using perspective views as input. %

\subsection{Metrics}

\paragraph{Validation Pair FID and FVD.} On validation sets we compare the set of generated frames to their paired real unseen frames in aggregate distribution. This evaluates both the quality and favors generations that adhere more closely to the true frames.
\vspace{-1em}
\paragraph{Reconstruction Metrics.} In the video-conditional setting, we directly compare generated frames to their real counterparts as is commonly done for evaluating novel view synthesis performance. We use PSNR, SSIM and LPIPS~\citep{lpips}. %
Note that evaluating reconstruction metrics in the conditional generative setting can be problematic as the desired output is inherently ambiguous. Namely, direct comparisons with the ground truth can favor mode covering solutions, that may be lower in diversity.
\vspace{-1em}
\paragraph{Clip Score (Clip).} We evaluate alignment to the supplied text prompt via clip score.
\vspace{-1em}
\paragraph{User Preference.} We additionally conduct a user study, where equirectangular video/images from our model and the baseline are shown side-by-side to the user along with the conditioning input and they are asked to select their preferred result. For this setting, we randomly subsample 20 videos for each comparison and conducted the study with 6 users.

\subsection{Text-conditional Generation}
We evaluate our model's ability to generate multi-view videos from a text prompt and compare it to 360DVD \citep{wang2024360dvd} that is our primary baseline. A quantitative comparison is summarized in \cref{tab:quant_text_to_video}. We note that our model outperforms 360DVD across all metrics. 
A side-by-side visual comparison is provided in \cref{fig:abl_vs_dvd360}, demonstrating that \ourmodel produces videos with higher image quality and sharper details. In contrast, 360DVD's outputs extremely blurry and undersaturated results that suffer from  insufficient warping near the top and bottom of the panoramas, hence leading to noticeable stretching artifacts when viewed in 3D, as we show in Appendix \cref{fig:compare_vs_dvd360_distort}. %

\input{figures_tex/tab_ours_vs_360dvd}

\subsection{Video-conditional Generation}
Our video-conditional model accepts both a single view video and a text prompt which can be obtained through captioning the input view.
During training, we randomly select one of the horizontal views, as shown in \cref{fig:emb},as the conditional one and do not apply any noise on it. We exclude conditioning on top and bottom views as this case is less common. 
During inference we directly treat the input video as one of our horizontal views.

For general videos, there are no existing models that consider the video-conditional panoramic video generation task. Therefore, we compare our model to existing image-conditional panorama image generation model, MVDiffusion, at the frame level. In particular, for our method, we first generate a 16 frame panorama video and then extract the middle frame.
We compare against the outpainting model from MVDiffusion and report the results in \cref{tab:ours_vs_mvdiffusion}. Since MVDiffusion does not cover the sky or ground regions, we only evaluate metrics on the 8 horizontal views. Our method scores significantly better on FID and reconstruction metrics, while being slightly worse on the clip score. Qualitatively we find that our method is much better at maintaining the style and scene scale/depth in the other generated views as demonstrated in the qualitative examples from \cref{fig:abl_vs_mvdiffusion}. 
We also tried comparing to PanoDiffusion but found that this model is prone to over-fitting to indoor room scenes. 
We additionally, perform video-conditional generation on out of distribution videos and show generated results in \cref{fig:teaser-top} and our project website.

\begin{table}[]
\centering
\begin{tabular}{l|rlrrr|l}
\toprule
                  & \multicolumn{5}{c}{Horizontal 8 views}  
                  & \multicolumn{1}{|c}{User}
                  \\
                  & \multicolumn{1}{l|}{FID $\downarrow$}  & \multicolumn{1}{l|}{Clip $\uparrow$}            & \multicolumn{1}{l|}{PSNR $\uparrow$}          & \multicolumn{1}{l|}{LPIPS $\downarrow$}          
                  & \multicolumn{1}{l}{SSIM $\uparrow$} 
                  & \multicolumn{1}{|l}{Pref$\uparrow$} 
                  \\ \midrule
MVDiffusion       & \multicolumn{1}{r|}{96.8}                    & \multicolumn{1}{r|}{\textbf{29.7}} & \multicolumn{1}{r|}{13.4}          & \multicolumn{1}{r|}{0.568}          
    & \multicolumn{1}{r|}{0.485}
    & 23\%
    \\
Ours (multi-task) & \multicolumn{1}{r|}{\textbf{63.2}}       & \multicolumn{1}{r|}{28.5}          & \multicolumn{1}{r|}{\textbf{17.6}} & \multicolumn{1}{r|}{\textbf{0.457}} & \textbf{0.636}           
    & \textbf{77\%}                   
\\ \midrule
\end{tabular}
\caption{Quantitative comparison of single view video-conditional panorama generation with image panorama outpainting method MVDiffusion. We extract the middle frame from our 16 frame generations to compare at a per image level. 
}
\label{tab:ours_vs_mvdiffusion}
\end{table}

\subsection{Autoregressive Generation}
To demonstrate our model's performance on long video generation, we run 4 iterations of autoregression, resulting in a total of $4 \times 15 + 1 = 61$ frames for the panorama videos. We observe that, despite using noise augmentation, autoregressive errors gradually accumulate, causing the scene to become blurry. To mitigate this, the noise-augmentation value can be increased during inference to regenerate finer details, though this introduces slight flickering due to the newly added details. Ideally, a dynamic system could be developed to increase the value when blurriness occurs and reduce it otherwise, minimizing flickering while keeping pixel quality high—an avenue we leave for future work. We provide examples of  extracted frames from our autoregressively generated videos in \cref{fig:teaser-top} and \cref{fig:appendix_autoregressive}. Please see our website for best viewing of long video generations.

\subsection{Ablations}
We ablate the main components of our method and include additinal ablations on shifting the noise schedule of the base model, the architecture for conditioning on image frames and noise augmentation in \cref{sec:app_additional_ablations}.

\input{figures_tex/tab_videocond_ablation}

\paragraph{Random Matrix vs. Fixed Matrix.}
\input{figures_tex/ablation_full_vs_random_matrix}
During training, we can fit a maximum of 6 time frames with 8 multi-views in memory. However, at inference we wish to generate 16 frames which is the native frame length for our base video model and aligns with 360DVD.
To enable this we employ the randomized matrix strategy described in \cref{method:autoregressive}.
To evaluate the benefit of this strategy, we compare $8\times 16$ video-conditional generations from a model that was trained with the ``random-matrix'' strategy using an even mix of $8\times 6$, $4\times 12$ and $3\times16$ with one using a ``fixed-matrix'' strategy trained with only the $8\times 6$ setting. We include a quantitative comparison in \cref{tab:quant_video_cond} and qualitative examples in \cref{fig:abl_full_vs_random_matrix}.
From the comparison, we see that the fixed-matrix trained model can create more blurry regions in its generations which is also reflected in significantly higher FID and FVD and somewhat lower clip score. 
Reconstruction metrics are very similar but slightly prefer the fixed-matrix model. We hypothesize that focusing training more on the 8 view case could slightly improve the global color consistency at the cost of worse visual quality.

\paragraph{Multi-task Training.}
We find that we can train one unified model to handle text-only conditioning, single view video conditioning and autoregressive conditioning.
In \cref{tab:quant_video_cond} we also quantitatively compare our multi-task model with one only trained for the video conditional setting.
For all the metrics, the multi-task model is marginally worse but very close indicating that we can train our model jointly with negligible impact to the quality.
We also observe on some OOD conditions, that the random conditioned model tends to improve pixel quality slightly as seen in \cref{fig:abl_fixed_vs_rand_cond} which could be due to better generalization from multi-task training. 
\input{figures_tex/ablation_fixed_vs_rand_conv}

%% file: figures_tex/ours_vs_dvd360.tex
\begin{figure}
\centering
\includegraphics[width=0.9\textwidth]{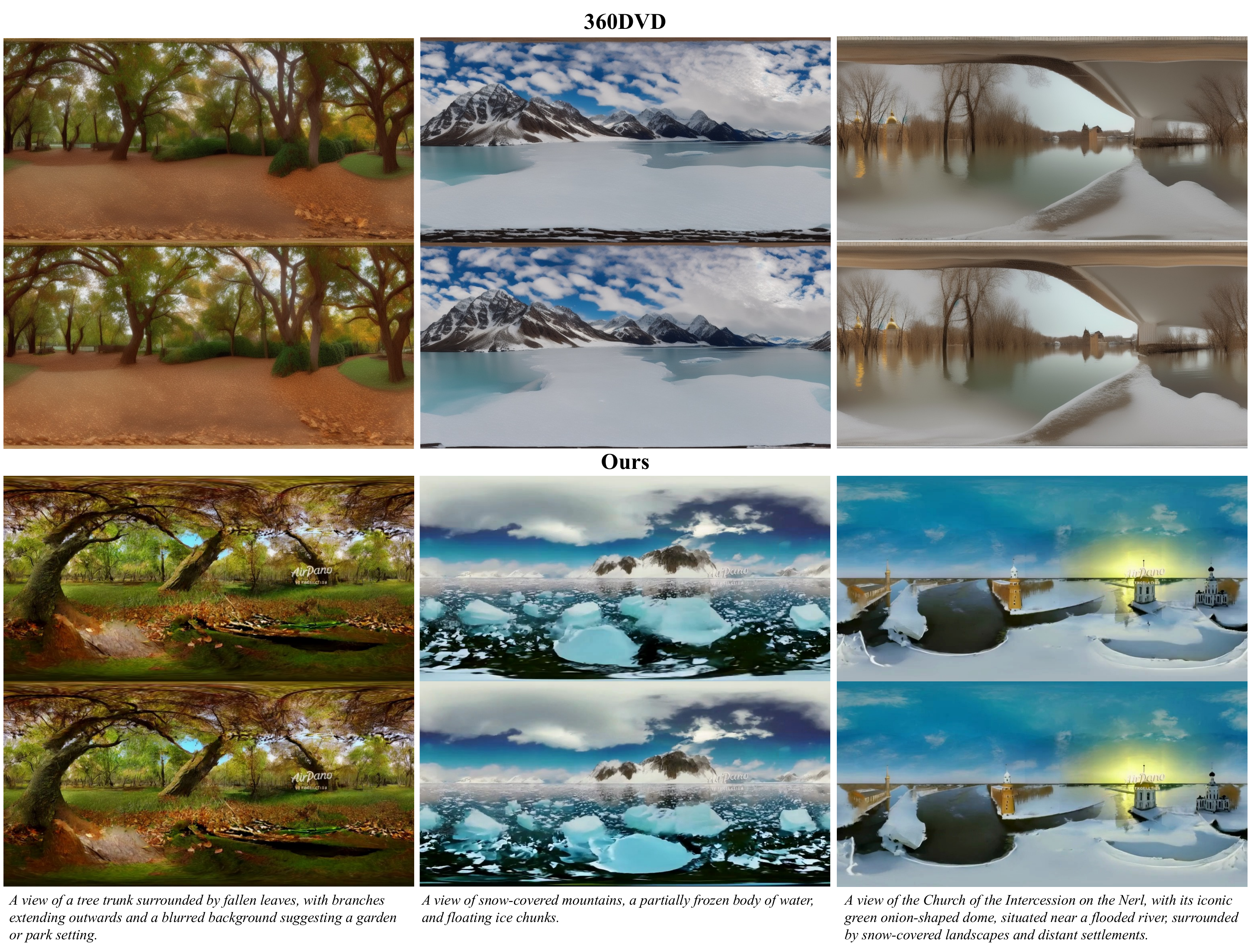}    
\caption{Qualitative figure compare text conditional video generation, 360DVD VS ours. The pixel quality of 360DVD is lower and distortion near the poles (top and bottom) is worse. }
\label{fig:abl_vs_dvd360}
\end{figure}

%% file: figures_tex/ours_vs_mvdiffusion.tex
\begin{figure}
\centering
\includegraphics[width=0.9\textwidth]{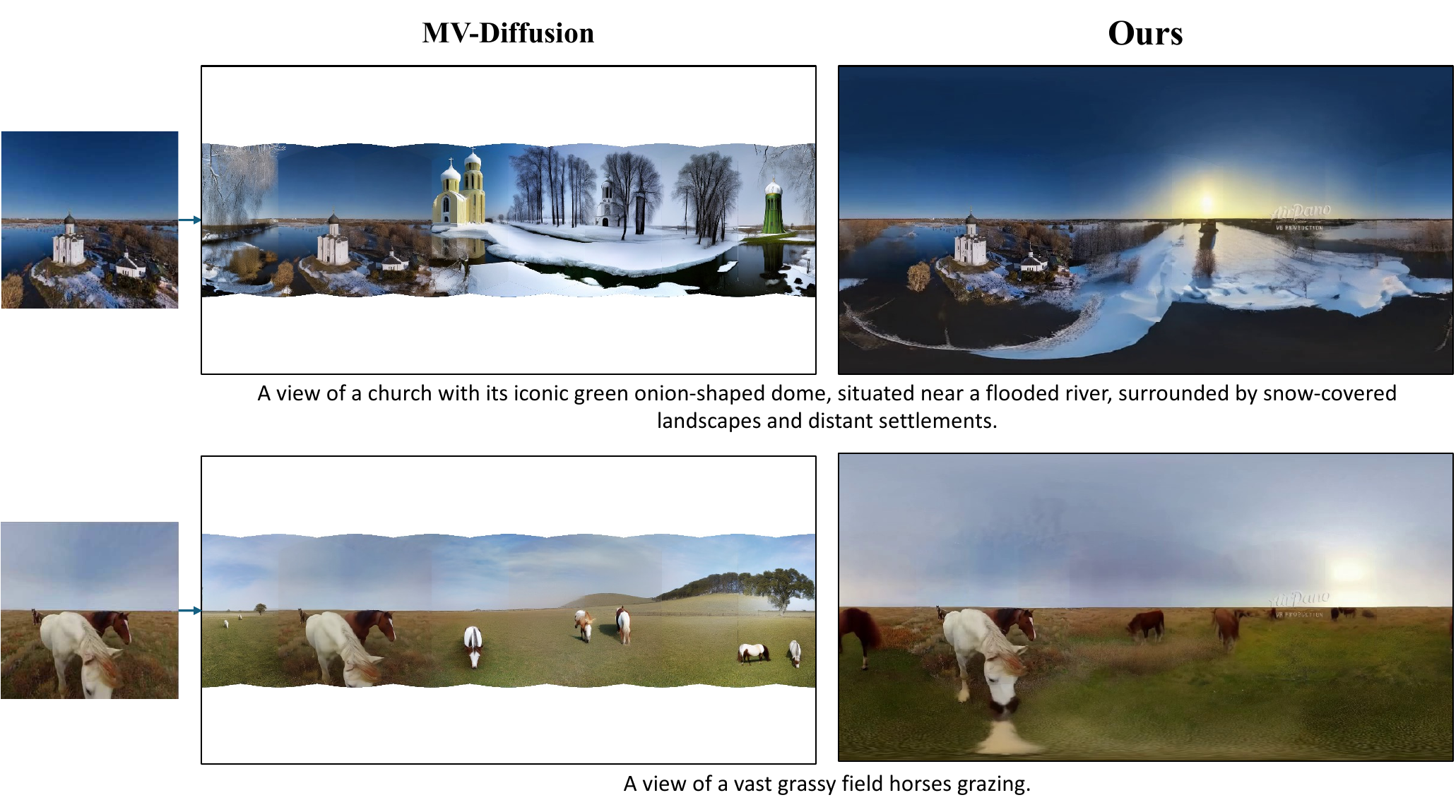}    
\caption{Qualitative figure comparing video conditional generation, MVDiffusion VS ours. Note that MVDiffusion can only outpaint each frame of the video separately. MVDiffusion is worse at maintaining the structure and style of the input view globally compared to ours. For example the sky color and the scales and depths of objects is less consistent for MVDiffusion.}
\label{fig:abl_vs_mvdiffusion}
\end{figure}

%% file: figures_tex/tab_ours_vs_360dvd.tex
\begin{table}[!thbp]
\centering
\begin{tabular}{l|lrr|lrr|l}
\toprule
& \multicolumn{3}{c|}{Panorama}                               
& \multicolumn{3}{c|}{Horizontal 8 views}                         
& \multicolumn{1}{c}{User}
\\
 &
  \multicolumn{1}{l|}{FID\textsubscript{pair} $\downarrow$} &
  \multicolumn{1}{l|}{FVD\textsubscript{pair} $\downarrow$} &
  \multicolumn{1}{l|}{Clip $\uparrow$} &
  \multicolumn{1}{l|}{FID\textsubscript{pair} $\downarrow$} &
  \multicolumn{1}{l|}{FVD\textsubscript{pair} $\downarrow$} &
  \multicolumn{1}{l|}{Clip $\uparrow$} &
  \multicolumn{1}{c}{Pref$\uparrow$} \\ 
  \midrule
360DVD & \multicolumn{1}{r|}{160} & \multicolumn{1}{r|}{1942} & 28.4 & \multicolumn{1}{r|}{128.7} & \multicolumn{1}{r|}{958.2} & 27.6 &
28\%
\\
Ours (multi-task) &
  \multicolumn{1}{r|}{\textbf{136}} &
  \multicolumn{1}{r|}{\textbf{1258}} &
  \textbf{29.8} &
  \multicolumn{1}{r|}{\textbf{91.3}} &
  \multicolumn{1}{r|}{\textbf{600.5}} &
  \textbf{28.9}  &
  \textbf{72\%}
\\ \bottomrule
\end{tabular}

\caption{Quantitative comparison for text-conditional panorama video generation.} %
\label{tab:quant_text_to_video}
\end{table}

%% file: figures_tex/tab_videocond_ablation.tex
\begin{table}[]
\centering
\resizebox{\textwidth}{!}{
\begin{tabular}{lc|rrrr|rrrr}
\toprule
\multicolumn{2}{c|}{Ours Ablation} &
  \multicolumn{4}{c|}{Panorama} &
  \multicolumn{4}{c}{Horizontal 8 view videos} \\
\multicolumn{1}{c|}{multi-task} &
  rand-mat &
  \multicolumn{1}{l|}{FID$\downarrow$} %
  & \multicolumn{1}{l|}{FVD$\downarrow$} %
  & \multicolumn{1}{l|}{Clip $\uparrow$} &
  \multicolumn{1}{l|}{PSNR $\uparrow$} &
  \multicolumn{1}{l|}{FID $\downarrow$} &
  \multicolumn{1}{l|}{FVD $\downarrow$} &
  \multicolumn{1}{l|}{Clip $\uparrow$} &
  \multicolumn{1}{l}{PSNR $\uparrow$} \\ \midrule
\multicolumn{1}{c|}{$\checkmark$} &
  $\checkmark$ &
  \multicolumn{1}{r|}{\textbf{98}} &
  \multicolumn{1}{r|}{916} &
  \multicolumn{1}{r|}{\textbf{29.6}} &
  15.9 &
  \multicolumn{1}{r|}{49.8} &
  \multicolumn{1}{r|}{258} &
  \multicolumn{1}{r|}{\textbf{28.6}} &
  17.6 \\
\multicolumn{1}{c|}{$\times$} &
  $\checkmark$ &
  \multicolumn{1}{r|}{103} &
  \multicolumn{1}{r|}{\textbf{861}} &
  \multicolumn{1}{r|}{28.9} &
  16.0 &
  \multicolumn{1}{r|}{\textbf{48.4}} &
  \multicolumn{1}{r|}{\textbf{255}} &
  \multicolumn{1}{r|}{28.2} &
  17.3 \\
\multicolumn{1}{c|}{$\times$} &
  $\times$ &
  \multicolumn{1}{r|}{124} &
  \multicolumn{1}{r|}{999} &
  \multicolumn{1}{r|}{27.1} &
  \textbf{17.0} &
  \multicolumn{1}{r|}{69.8} &
  \multicolumn{1}{r|}{445} &
  \multicolumn{1}{r|}{26.0} &
  \textbf{18.5}
\\ \bottomrule
\end{tabular}
}
\caption{Quantitative ablations of our model on single view video-conditional panoramic video generation. Training our model to be multi-task capable incurs a negligible drop in performance. Randomizing the matrix of frames during training results in much improved video quality at a slightly worse color consistency as measured by PSNR. 
}
\label{tab:quant_video_cond}
\end{table}

%% file: figures_tex/ablation_full_vs_random_matrix.tex
\begin{figure}
\centering
\includegraphics[width=0.9\textwidth]{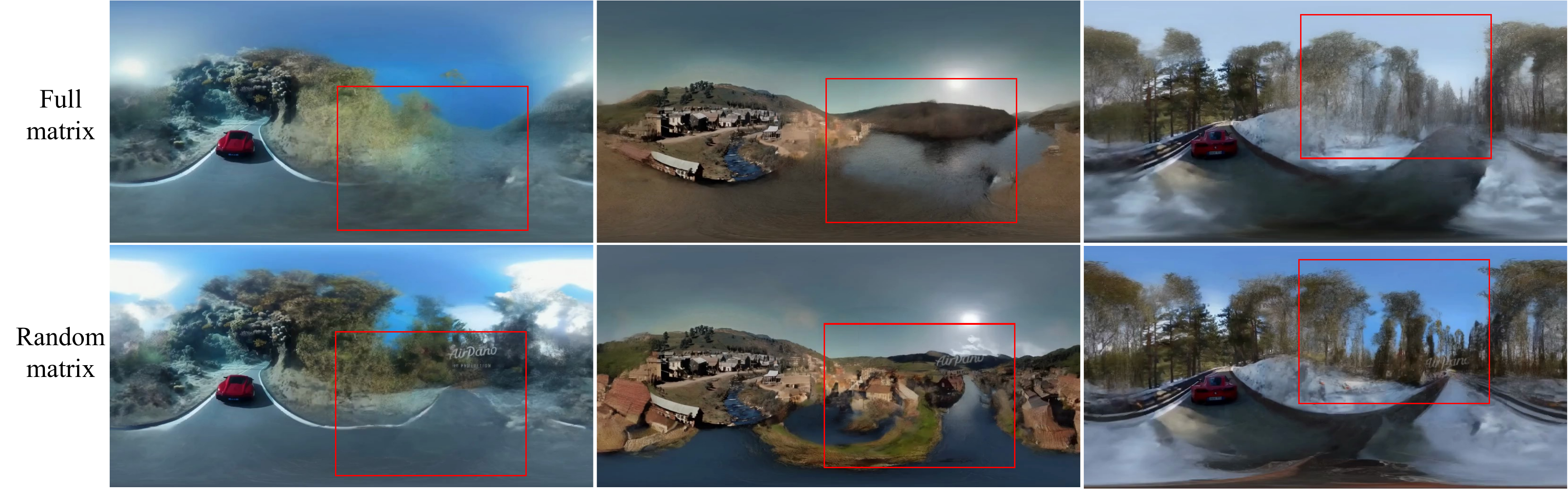}
\caption{Qualitative figure comparing full matrix and random matrix training. Random matrix training generates more high frequency details.}
\label{fig:abl_full_vs_random_matrix}
\end{figure}

%% file: figures_tex/ablation_fixed_vs_rand_conv.tex
\begin{figure}
\centering
\includegraphics[width=\textwidth]{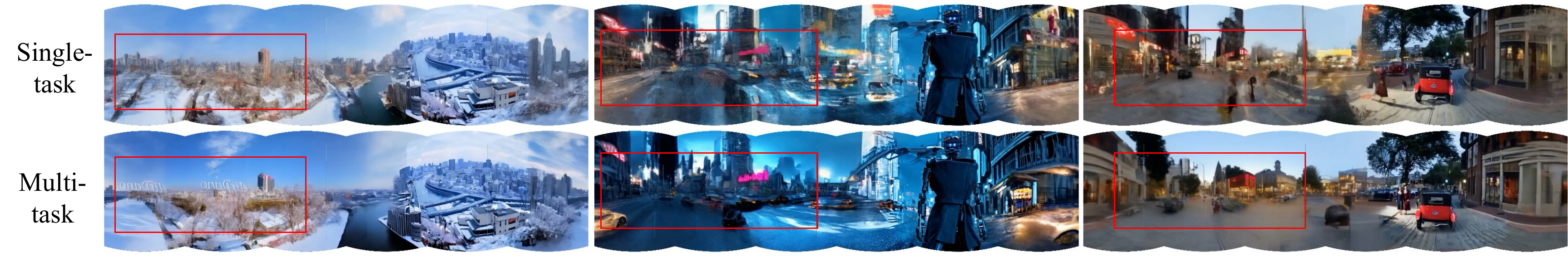}

\caption{Qualitative figure comparing our single task vs multi-task model both generating 6 views on out of distribution video. Multi-task training provides better pixel quality. Moreover, with multi-task training, we can train one unified model for different tasks including video conditional generation and auto-regressive generation. }
\label{fig:abl_fixed_vs_rand_cond}
\end{figure}

%% file: sections/conclusion.tex
\section{Conclusion}
We present \ourmodel, a model for panoramaic video generation. \ourmodel augments a pretrained video diffusion model with the ability to generate consistent multiview videos that together cover a full panoramic video. We train \ourmodel in a unified manner with flexible conditioning supporting text and single-view video-conditioning and further support auto-regressive generation of longer videos.

Although \ourmodel demonstrates compelling results, there is still room for further improvement. The generation capabilities of our model are restricted by the performance of the base video model and further improvements could be obtained by applying these techniques to more powerful video diffusion models. 
Our model currently requires the field of view and elevation of the conditioning input to be sufficiently close to the configuration used in training. This could be addressed by estimating these parameters as demonstrated by recent work in the image generation domain~\citep{yuan2024camfreediff}.
Our autoregressive generation balances a trade-off between maintaining image quality over time and consistency between windows which motivates investigating methods that could efficiently achieve both. 

\paragraph{Acknowledgments} We would like to thank Frank Shen, Xuanchi Ren, Ruofan Liang, Zian Wang and Jun Gao for their valuable contributions to training code infrastructure that was used for training VideoPanda with different base model.

%% file: sections/appendix.tex
\newpage

\section{Additional Ablations}
\renewcommand{\thefigure}{A\arabic{figure}} %
\setcounter{figure}{0}
\label{sec:app_additional_ablations}

\subsection{IP Adapter VS Cat}
A popular method for image-conditioning in image diffusion models is IP Adapter~\cite{ye2023ip-adapter}. A CNN feature extractor takes the conditional input views and extracts features that will then be added to the intermediate features in the diffusion model forward pass. Here we compare it to using the conditioning method from CAT3D that directly uses conditional inputs as frames to the diffusion model without noising.
We generally find that they are similar but IP adapter can exhibit more abrupt transitions between the input condition and neighbouring regions in the generated panorama. We show a few examples in \cref{fig:abl_cat_vs_ip}.

\input{figures_tex/ablation_cat_vs_ip}
\subsection{Ablating the effects of shifting the noise schedule} \label{appendix:ablating_shiftsnr}

During inference we use up to $8\times 16=128$ frames which is much larger than the $16$ frames used by the base video model.
As mentioned in \cref{method:model_design} the increased data dimensionality also requires a corresponding increase in terminal noise to minimize the terminal step gap with the noise prior. 
In particular we interpolate between the standard noise schedule and a noise schedule that has been shifted by $10$.
We compare these qualitatively in  \cref{fig:abl_eps_vs_vpredsnr}. Note that without changing the noise schedule, the model is largely incapable of generating plain regions such as clear sky or white snow fields and instead fills in the frame with visual clutter.

\input{figures_tex/ablation_eps_vs_vpredinterpSNR}

\subsection{Ablating the effect of noise augmentation for autoregressive generation} \label{appendix:ablating_noise_aug}
In this section, we qualitatively analyze the impact of noise augmentation during training on the model's autoregressive generation performance. To demonstrate this, we compare two models: one trained with noise augmentation and the other without. To maximize the effect of error accumulation, we use both models to 6 frames at a time, for a total of 10 iterations to get a video consisting of $10 \times 5 + 1 = 51$ frames. \Cref{fig:abl_noise_aug} shows a side-by-side comparison of the different scenarios.

As autoregressive iterations increase, the model without noise-augmentation produces increasingly saturated frames. While the model trained with noise augmentation also shows some degradation, it maintains significantly better output quality over time, demonstrating its usefulness in reducing the severity of error accumulation.

\input{figures_tex/ablation_noise_aug}

\subsection{Ablating the effects of freezing base model layers} 
\input{figures_tex/ablation_freeze_text}
When finetuning our model for multi-view generation we choose to freeze the base model layers. We ablate this choice qualitatively here on the text conditional panorama video generation task. We evaluate out of distribution prompts that make the overfitting behaviour very obvious when not freezing any base layers as can be seen in Fig.\ref{fig:abl_wo_freeze}.

\input{figures_tex/ours_vs_360dvd_distort}

\section{Additional Training Details}
\renewcommand{\thefigure}{B\arabic{figure}} %
\renewcommand{\thetable}{B\arabic{table}} %
\setcounter{figure}{0}
\setcounter{table}{0}
\label{appendix:extra_train_detail}
During the first stage of training we adapt the base video model towards the shifted and interpolated noise schedule as well as the v-prediction parameterization. This stage is trained for $10,000$ iterations on the original dataset and a batch size of $128$.
Following that we insert the multi-view attention layers and train our model using the multiview video data.
The batch size for this phase is $32$ and we train these models for $15,000$ iterations.
Both stages use a constant learning rate of 
$0.0001$.
Most of our experiments are conducted on 32 A100 GPUs (or lower using gradient accumulation).

\section{Evaluation on Out of Distribution Prompts}
\renewcommand{\thefigure}{C\arabic{figure}} %
\renewcommand{\thetable}{C\arabic{table}} %
\setcounter{figure}{0}
\setcounter{table}{0}
\label{appendix:ood_eval}
\input{figures_tex/tab_quant_ood_text.tex}
For evaluating our model on out of distribution prompts for text-conditional video generation, we use the same inference setting as before and tabulate the results in \cref{tab:quant_ood_text}.
We use prompts from VBench, in particular all 946 prompts from the ``all-dimensions category".
For each prompt we sample 3 different videos.
As we lack ground truth videos we cannot compute pairwise FVD.
Instead, we evaluated non-paired FID and FVD for the OOD text-conditional case, using the popular video dataset HDVila for the reference set. 
In particular, we use 3,000 random videos from HDVila for FVD computations and use the first frames from the same set for FID computations. 
As this reference set consists primarily of perspective view videos, we only evaluate extracted perspective views from our generated panorama videos.
For perspective view extraction, we also included views with non-zero elevation. Specifically, we create an additional setting where we extract 8 views in total with 4 views at negative 60 degree elevation looking downwards and another 4 views at positive 60 degree elevation looking upwards. The FOV is kept at 90 degrees. We refer to this setting as ``Elevation=+/-60degree Views''. These views better capture a complete picture of the panorama while still remaining within the distribution of natural camera angles.

VideoPanda significantly outperforms 360DVD in the 60 degree elevation views, highlighting its superior ability to generate the ground and sky views, which are distorted in the equirectangular representation used by 360DVD.

\section{Different Base Video Model}
\renewcommand{\thefigure}{D\arabic{figure}} %
\renewcommand{\thetable}{D\arabic{table}} %
\setcounter{figure}{0}
\setcounter{table}{0}
\label{appendix:ood_eval}
\label{sec:app_different_base_video_model}
Our VideoPanda method can be flexibly applied to other base models with different architectures.
To demonstrate this, we apply our framework on top of the open source CogVideoX-2B video diffusion transformer model. 
Analogous to before, we leave the standard 3D video attention to process each of the video views independently and interleave them with additional per-frame multiview attention layers.
We find that the model benefits from the increased capabilities of the base video model and greatly improves in visual details and has some improvements to the semantic scene coherence as well.
We include some side-by-side comparisons of generated videos using the CogVideoX-2B base model in Fig.~\ref{fig:appendix_base_model_comp}.
We also apply the random matrix method to extend the test-time temporal window. For 8 views our model training can only fit 5 temporal tokens corresponding to 17 frames, however at 3 views we can fit the full 49 frames which is the native number of frames for the base model and we include all the combinations between these two settings. Additionally, the superior performance gained by using the CogVideoX-2B base model with VideoPanda is clearly seen by the quantitative evaluation in Tab.~\ref{tab:quant_ood_text} above.

\begin{figure}
\centering
\includegraphics[width=0.9\textwidth]{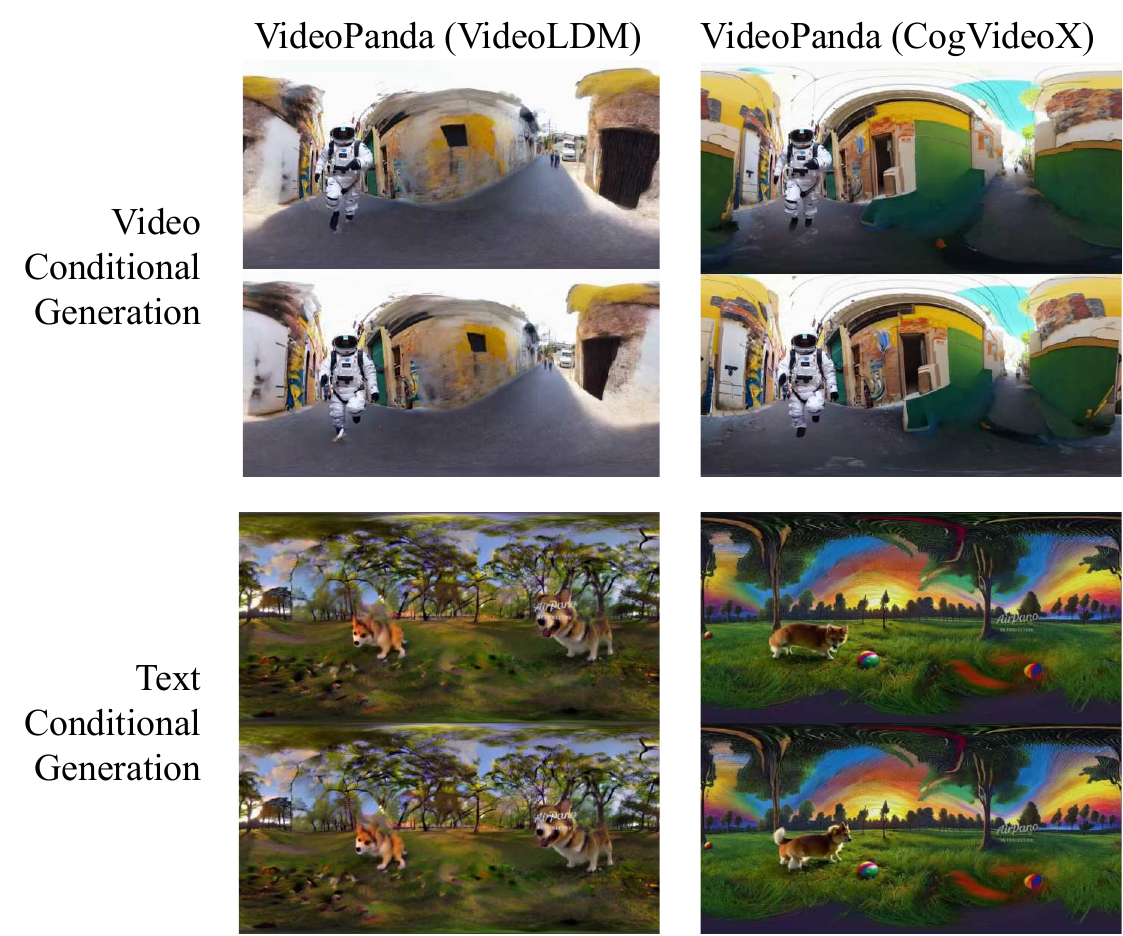}    
\caption{Qualitative comparison of generated videos from different base model. In the left column is VideoPanda using the VideoLDM base model and in the right column it is using CogVideoX-2B as the base model. In the top row, we are showing a single-view video conditional generation result and in the bottom row, a text-conditional video generation. We see that using the CogVideoX base model helps especially in OOD text-conditional generation settings where it shows stronger generalization.}
\label{fig:appendix_base_model_comp}
\end{figure}

\section{Additional Video Conditional Results}
\renewcommand{\thefigure}{E\arabic{figure}} %
\renewcommand{\thetable}{E\arabic{table}} %
\setcounter{figure}{0}
\setcounter{table}{0}
We show more video conditional generation results in Fig.~\ref{fig:appendix_autoregressive} where we also apply autoregressive generation to extend the video length.

\begin{figure}[h!]
\centering
\includegraphics[width=0.9\textwidth]{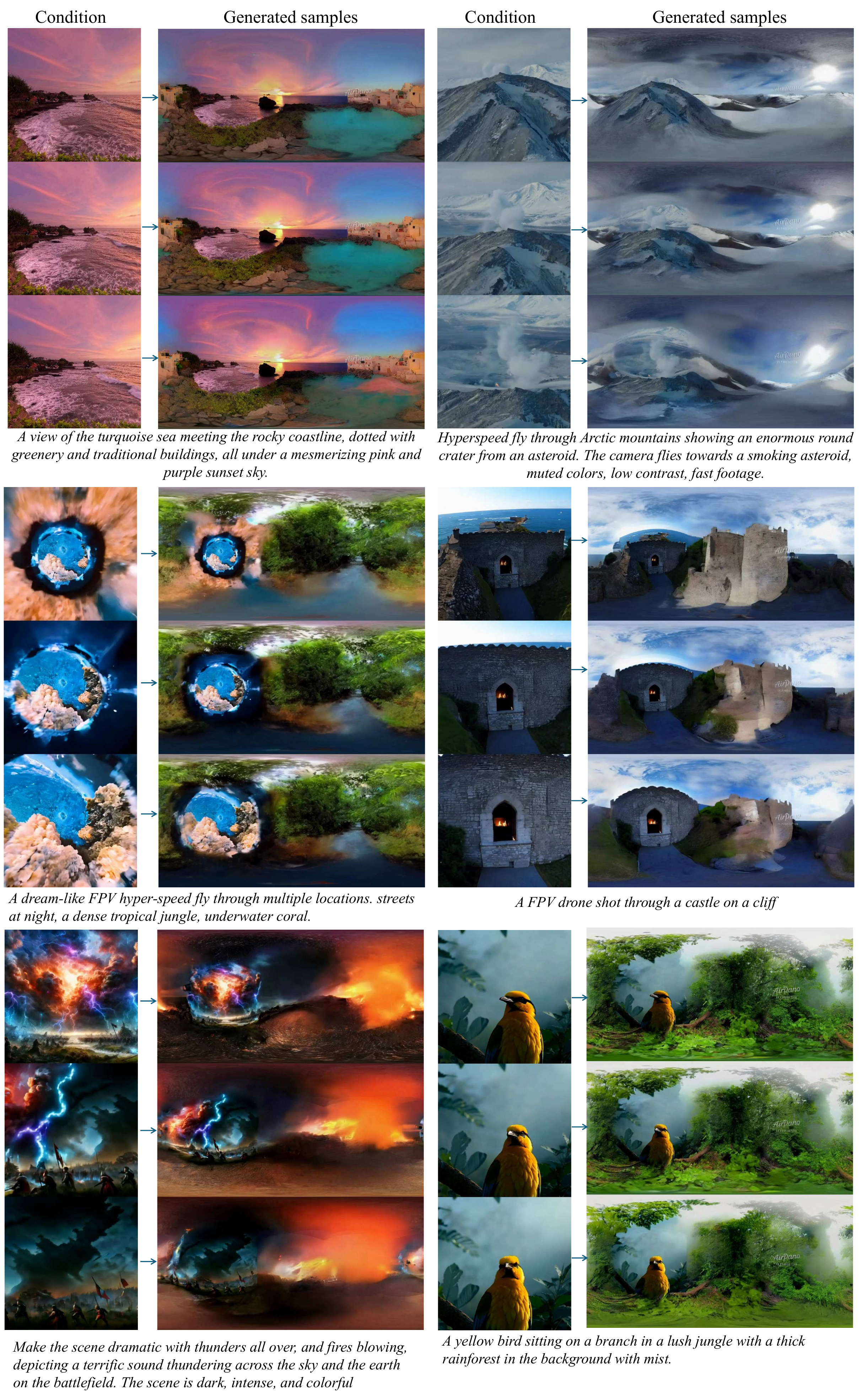}    
\caption{More results of video conditional autoregressive generation on out of distribution videos.}
\label{fig:appendix_autoregressive}
\end{figure}

%% file: figures_tex/ablation_cat_vs_ip.tex
\begin{figure}[h!]
\centering
\includegraphics[width=\textwidth]{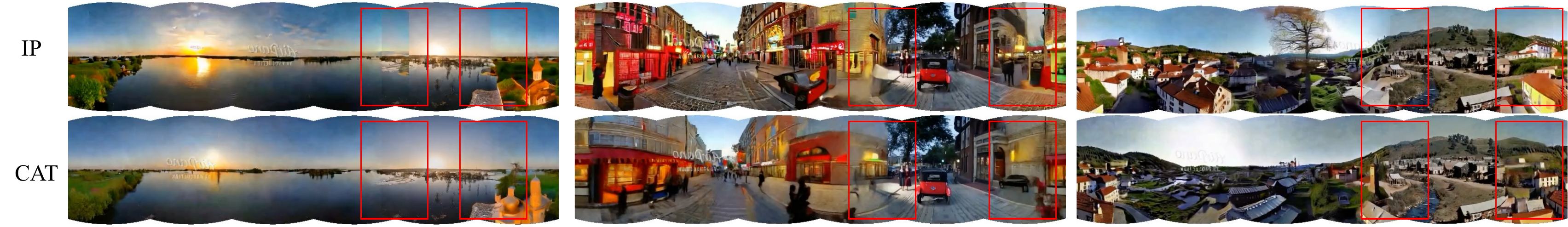}
\caption{Qualitative figure comparing IP vs CAT type architecture for input conditioning. When using IP adapter, the consistency between input conditioning views and neighbouring views (highlighted in red box) is worse compare to CAT. }
\label{fig:abl_cat_vs_ip}
\end{figure}

%% file: figures_tex/ablation_eps_vs_vpredinterpSNR.tex
\begin{figure}[h]
\centering
\includegraphics[width=\textwidth]{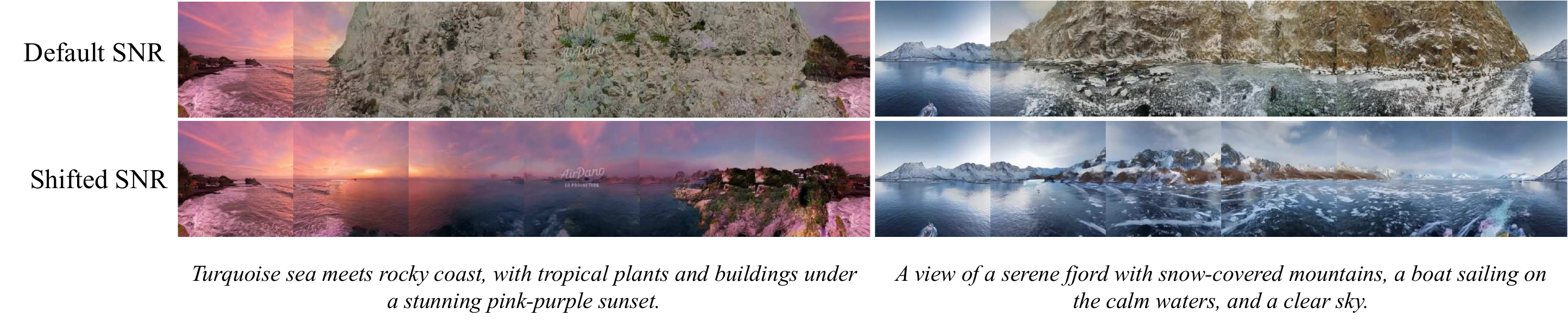}
\caption{
Qualitative comparison of shifting the noise schedule in the video-conditioned setting. Each of the six horizontal views is visualized independently before stitching into a panorama. Without shifting toward higher noise levels, the model struggles to generate clear skies or water, introducing objects that disrupt scene cohesion (e.g., sudden mountains and rocks).
}
\label{fig:abl_eps_vs_vpredsnr}
\end{figure}

%% file: figures_tex/ablation_noise_aug.tex
\begin{figure}[h!]
\centering
\includegraphics[width=\textwidth]{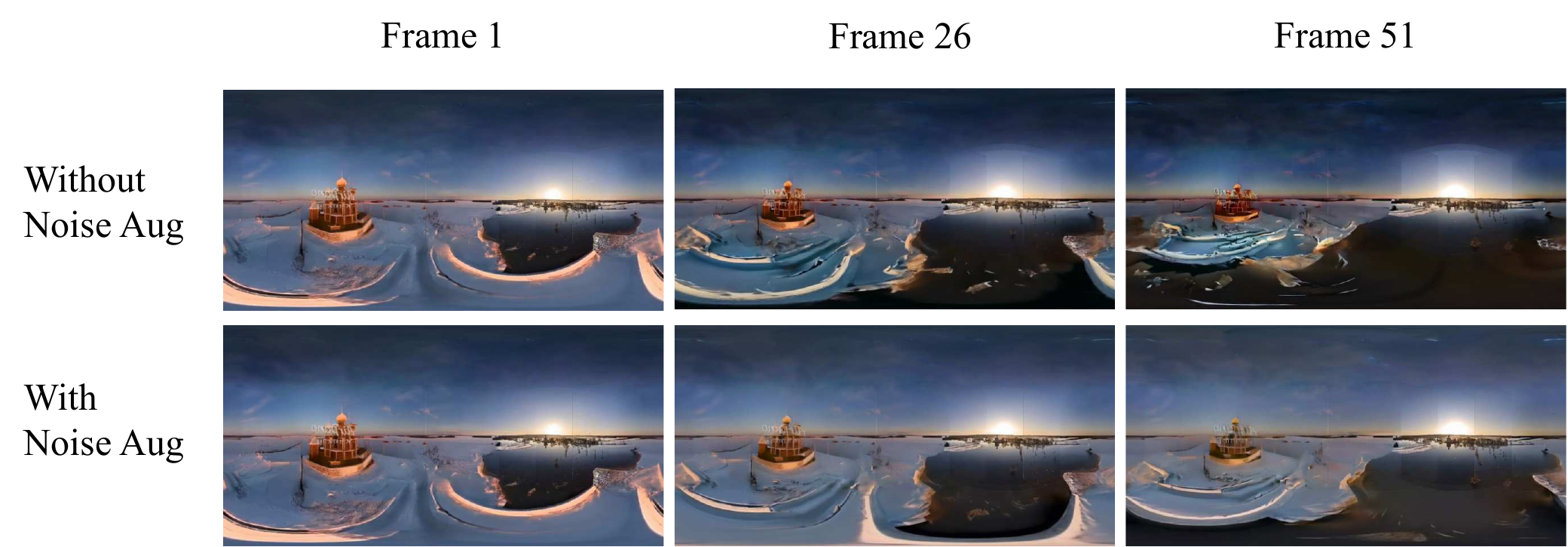}
\caption{Qualitative comparison of autoregressive generation with and without noise augmentation. Both models exhibit a decline in output quality over time, but the model trained without noise augmentation shows a more rapid and severe degradation, with frames becoming increasingly saturated. In contrast, the model with noise augmentation deteriorates more gradually.}
\label{fig:abl_noise_aug}
\end{figure}

%% file: figures_tex/ablation_freeze_text.tex
\begin{figure}[h!]
\centering
\includegraphics[width=\textwidth]{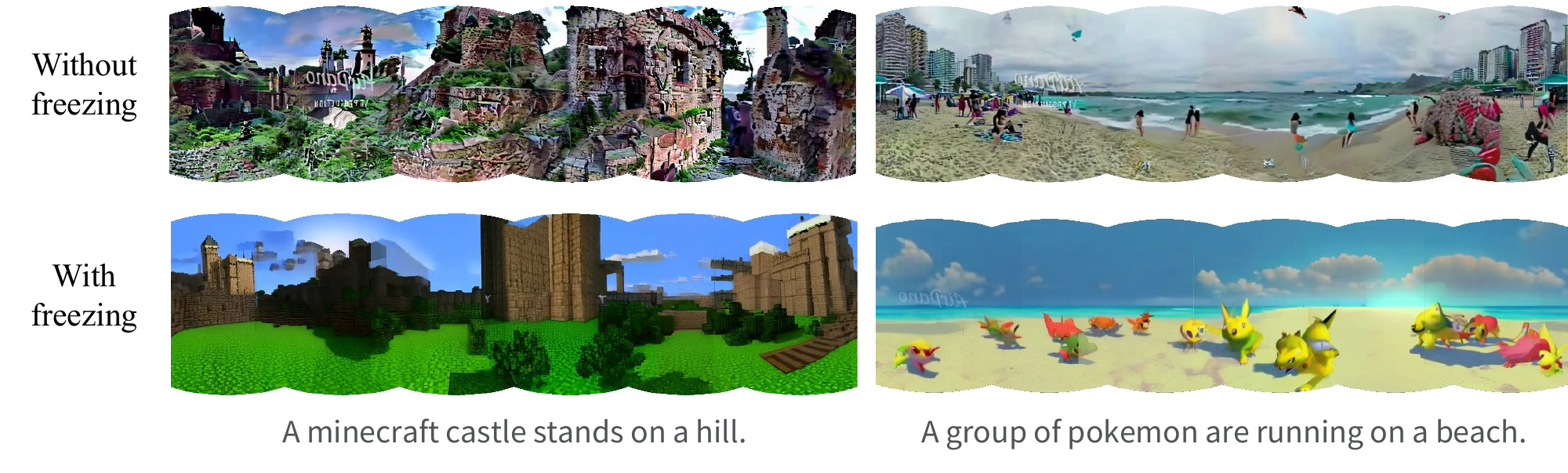}
\caption{Qualitative figure comparing text conditional panorama video generation using base model freezing vs no freezing. Freezing model weights is better able to retain some of the prior knowledge on out of distribution prompts. }
\label{fig:abl_wo_freeze}

\end{figure}

%% file: figures_tex/ours_vs_360dvd_distort.tex
\begin{figure}
\centering
\includegraphics[width=0.9\textwidth]{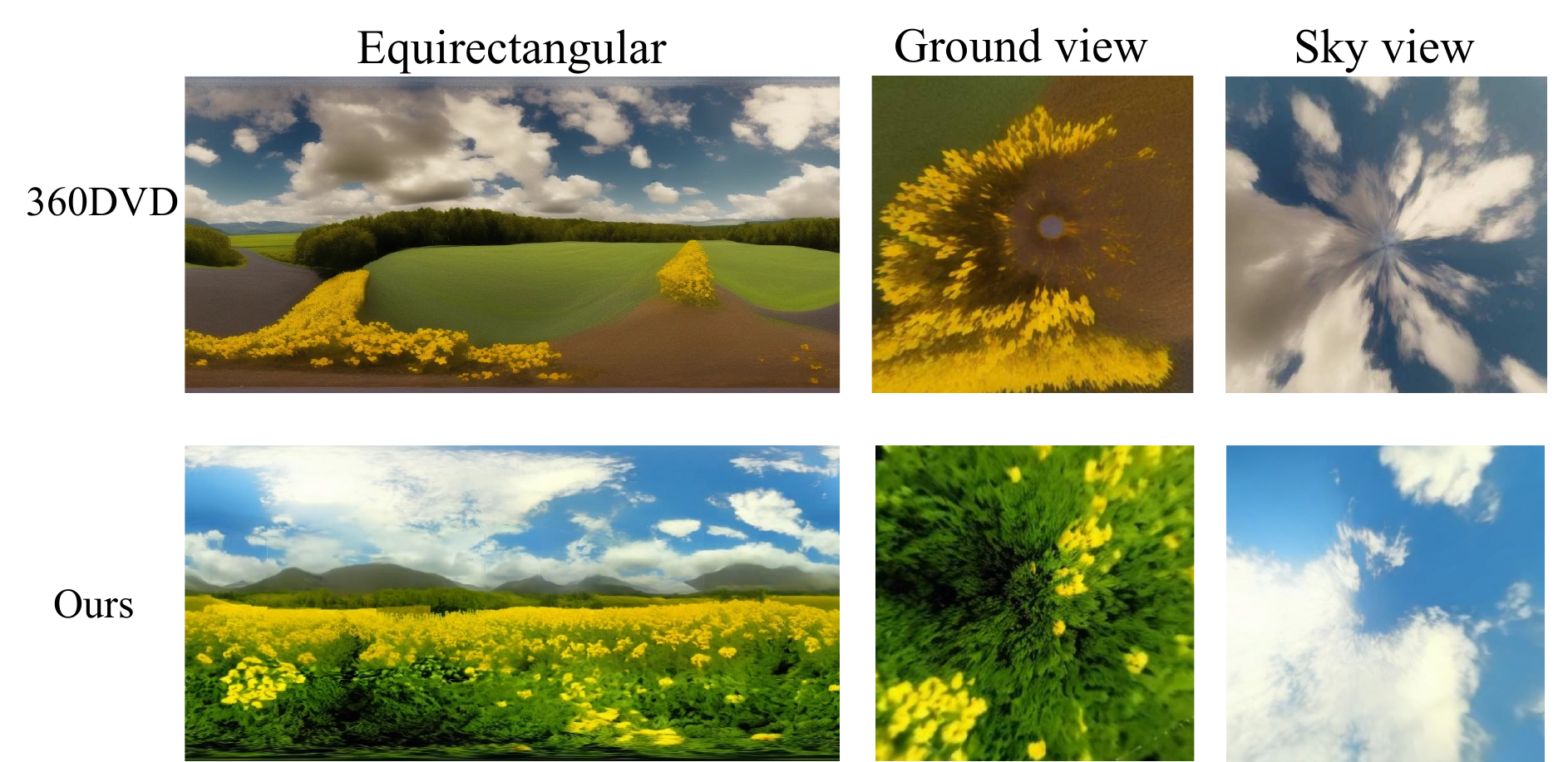}    
\caption{Qualitative figure comparing text conditional video generation, 360DVD VS ours and highlighting the distortion in 360DVD near the poles. Note that both generations were first transformed to the same equirectangular format before consistent sky and ground views were extracted. 360DVD struggles in these views as the distortion is highest here and deviates the most from perspective view images whereas we natively generate perspective views. }
\label{fig:compare_vs_dvd360_distort}
\end{figure}

%% file: figures_tex/tab_quant_ood_text.tex
\begin{table}[]
    \centering
    \resizebox{\textwidth}{!}{%
        \begin{tabular}{lc|ccc|ccc}
            \toprule
            \multicolumn{2}{c|}{Method} & \multicolumn{3}{c|}{Elevation=$\pm 60^\circ$ 8 Views} & \multicolumn{3}{c}{Horizontal 8 Views (elevation=0)} \\
            \midrule
            & & FID-COCO$\downarrow$ & FVD-W360$\downarrow$ & CS$\uparrow$ & FID-COCO$\downarrow$ & FVD-W360$\downarrow$ & CS$\uparrow$ \\
            \midrule
            \multicolumn{2}{l|}{360DVD} & $128.6$ & $901.1$ & $23.39$ & $\mathbf{91.8}$ & $801.9$ & $27.63$ \\
            \midrule
            \multirow{2}{*}{Ours} & (VideoLDM) & $\mathbf{115.0}$ & $\mathbf{826.6}$ & $\mathbf{24.12}$ & $92.6$ & $\mathbf{677.8}$ & $\mathbf{27.78}$ \\
            & (Cogvideo) & $\mathbf{\underline{93.4}}$ & $\mathbf{\underline{675.9}}$ & $\mathbf{\underline{25.99}}$ & $\mathbf{\underline{74.5}}$ & $\mathbf{\underline{624.7}}$ & $\mathbf{\underline{29.33}}$ \\
            \bottomrule
        \end{tabular}%
    }
    \caption{Quantitative evaluation of text conditional video generation on out of distribution prompts from Vbench All Dimensions (946 prompts). FID-COCO is FID to MS-COCO3k eval set and FVD-W360 is FVD to WEB360 training set.
    Note all compared methods use WEB360 as the training dataset.
    }
    \label{tab:quant_ood_text}
\end{table}

%% file: main.bbl
\begin{thebibliography}{69}
\providecommand{\natexlab}[1]{#1}
\providecommand{\url}[1]{\texttt{#1}}
\expandafter\ifx\csname urlstyle\endcsname\relax
  \providecommand{\doi}[1]{doi: #1}\else
  \providecommand{\doi}{doi: \begingroup \urlstyle{rm}\Url}\fi

\bibitem[Blattmann et~al.(2023{\natexlab{a}})Blattmann, Dockhorn, Kulal,
  Mendelevitch, Kilian, Lorenz, Levi, English, Voleti, Letts,
  et~al.]{blattmann2023stable}
Andreas Blattmann, Tim Dockhorn, Sumith Kulal, Daniel Mendelevitch, Maciej
  Kilian, Dominik Lorenz, Yam Levi, Zion English, Vikram Voleti, Adam Letts,
  et~al.
\newblock Stable video diffusion: Scaling latent video diffusion models to
  large datasets.
\newblock \emph{arXiv preprint arXiv:2311.15127}, 2023{\natexlab{a}}.

\bibitem[Blattmann et~al.(2023{\natexlab{b}})Blattmann, Rombach, Ling,
  Dockhorn, Kim, Fidler, and Kreis]{blattmann2023videoldm}
Andreas Blattmann, Robin Rombach, Huan Ling, Tim Dockhorn, Seung~Wook Kim,
  Sanja Fidler, and Karsten Kreis.
\newblock Align your latents: High-resolution video synthesis with latent
  diffusion models.
\newblock In \emph{IEEE Conference on Computer Vision and Pattern Recognition
  ({CVPR})}, 2023{\natexlab{b}}.

\bibitem[Brooks et~al.(2024)Brooks, Peebles, Holmes, DePue, Guo, Jing, Schnurr,
  Taylor, Luhman, Luhman, Ng, Wang, and Ramesh]{sora}
Tim Brooks, Bill Peebles, Connor Holmes, Will DePue, Yufei Guo, Li~Jing, David
  Schnurr, Joe Taylor, Troy Luhman, Eric Luhman, Clarence Ng, Ricky Wang, and
  Aditya Ramesh.
\newblock Video generation models as world simulators.
\newblock 2024.
\newblock URL
  \url{https://openai.com/research/video-generation-models-as-world-simulators}.

\bibitem[Chen et~al.(2024)Chen, Zhang, Cun, Xia, Wang, Weng, and
  Shan]{chen2024videocrafter2}
Haoxin Chen, Yong Zhang, Xiaodong Cun, Menghan Xia, Xintao Wang, Chao Weng, and
  Ying Shan.
\newblock Videocrafter2: Overcoming data limitations for high-quality video
  diffusion models, 2024.

\bibitem[Chen et~al.(2023)Chen, Xu, Ren, Cong, He, Xie, Sinha, Luo, Xiang, and
  Perez-Rua]{chen2023gentron}
Shoufa Chen, Mengmeng Xu, Jiawei Ren, Yuren Cong, Sen He, Yanping Xie, Animesh
  Sinha, Ping Luo, Tao Xiang, and Juan-Manuel Perez-Rua.
\newblock Gentron: Delving deep into diffusion transformers for image and video
  generation.
\newblock \emph{arXiv preprint arXiv:2312.04557}, 2023.

\bibitem[Gao* et~al.(2024)Gao*, Holynski*, Henzler, Brussee, Martin-Brualla,
  Srinivasan, Barron, and Poole*]{gao2024cat3d}
Ruiqi Gao*, Aleksander Holynski*, Philipp Henzler, Arthur Brussee, Ricardo
  Martin-Brualla, Pratul~P. Srinivasan, Jonathan~T. Barron, and Ben Poole*.
\newblock Cat3d: Create anything in 3d with multi-view diffusion models.
\newblock \emph{arXiv}, 2024.

\bibitem[Girdhar et~al.(2023)Girdhar, Singh, Brown, Duval, Azadi, Rambhatla,
  Shah, Yin, Parikh, and Misra]{girdhar2023emuvideo}
Rohit Girdhar, Mannat Singh, Andrew Brown, Quentin Duval, Samaneh Azadi,
  Sai~Saketh Rambhatla, Akbar Shah, Xi~Yin, Devi Parikh, and Ishan Misra.
\newblock Emu video: Factorizing text-to-video generation by explicit image
  conditioning.
\newblock \emph{arXiv preprint arXiv:2311.10709}, 2023.

\bibitem[Guo et~al.(2023)Guo, Yang, Rao, Liang, Wang, Qiao, Agrawala, Lin, and
  Dai]{guo2023animatediff}
Yuwei Guo, Ceyuan Yang, Anyi Rao, Zhengyang Liang, Yaohui Wang, Yu~Qiao,
  Maneesh Agrawala, Dahua Lin, and Bo~Dai.
\newblock Animatediff: Animate your personalized text-to-image diffusion models
  without specific tuning.
\newblock \emph{arXiv preprint arXiv:2307.04725}, 2023.

\bibitem[Gupta et~al.(2023)Gupta, Yu, Sohn, Gu, Hahn, Fei-Fei, Essa, Jiang, and
  Lezama]{gupta2023photorealistic}
Agrim Gupta, Lijun Yu, Kihyuk Sohn, Xiuye Gu, Meera Hahn, Li~Fei-Fei, Irfan
  Essa, Lu~Jiang, and Jos{\'e} Lezama.
\newblock Photorealistic video generation with diffusion models.
\newblock \emph{arXiv preprint arXiv:2312.06662}, 2023.

\bibitem[Hara \& Harada(2024)Hara and Harada]{hara2024magritte}
Takayuki Hara and Tatsuya Harada.
\newblock Magritte: Manipulative and generative 3d realization from image,
  topview and text.
\newblock \emph{arXiv preprint arXiv:2404.00345}, 2024.

\bibitem[Ho \& Salimans(2022)Ho and Salimans]{ho2022cfg}
Jonathan Ho and Tim Salimans.
\newblock Classifier-free diffusion guidance.
\newblock \emph{arXiv preprint arXiv:2207.12598}, 2022.

\bibitem[Ho et~al.(2020)Ho, Jain, and Abbeel]{ho2020denoising}
Jonathan Ho, Ajay Jain, and Pieter Abbeel.
\newblock Denoising diffusion probabilistic models.
\newblock \emph{Advances in neural information processing systems},
  33:\penalty0 6840--6851, 2020.

\bibitem[Ho et~al.(2021)Ho, Saharia, Chan, Fleet, Norouzi, and
  Salimans]{ho2021cascaded}
Jonathan Ho, Chitwan Saharia, William Chan, David~J Fleet, Mohammad Norouzi,
  and Tim Salimans.
\newblock Cascaded diffusion models for high fidelity image generation.
\newblock \emph{arXiv preprint arXiv:2106.15282}, 2021.

\bibitem[Ho et~al.(2022)Ho, Chan, Saharia, Whang, Gao, Gritsenko, Kingma,
  Poole, Norouzi, Fleet, et~al.]{ho2022imagen}
Jonathan Ho, William Chan, Chitwan Saharia, Jay Whang, Ruiqi Gao, Alexey
  Gritsenko, Diederik~P Kingma, Ben Poole, Mohammad Norouzi, David~J Fleet,
  et~al.
\newblock Imagen video: High definition video generation with diffusion models.
\newblock \emph{arXiv preprint arXiv:2210.02303}, 2022.

\bibitem[Hong et~al.(2022)Hong, Ding, Zheng, Liu, and Tang]{hong2022cogvideo}
Wenyi Hong, Ming Ding, Wendi Zheng, Xinghan Liu, and Jie Tang.
\newblock Cogvideo: Large-scale pretraining for text-to-video generation via
  transformers.
\newblock \emph{arXiv preprint arXiv:2205.15868}, 2022.

\bibitem[Hoogeboom et~al.(2023)Hoogeboom, Heek, and
  Salimans]{Hoogeboom2023simplediffusion}
Emiel Hoogeboom, Jonathan Heek, and Tim Salimans.
\newblock simple diffusion: End-to-end diffusion for high resolution images.
\newblock In \emph{International Conference on Machine Learning}, 2023.
\newblock URL \url{https://api.semanticscholar.org/CorpusID:256274516}.

\bibitem[Hu et~al.(2024)Hu, Zhou, Jampani, and Tulsiani]{hu2024mvdfusion}
Hanzhe Hu, Zhizhuo Zhou, Varun Jampani, and Shubham Tulsiani.
\newblock Mvd-fusion: Single-view 3d via depth-consistent multi-view
  generation.
\newblock In \emph{CVPR}, 2024.

\bibitem[Huang et~al.(2023)Huang, Wen, Dong, Wang, Li, Chen, Cao, Liang, Qiao,
  Dai, et~al.]{huang2023epidiff}
Zehuan Huang, Hao Wen, Junting Dong, Yaohui Wang, Yangguang Li, Xinyuan Chen,
  Yan-Pei Cao, Ding Liang, Yu~Qiao, Bo~Dai, et~al.
\newblock Epidiff: Enhancing multi-view synthesis via localized
  epipolar-constrained diffusion.
\newblock \emph{arXiv preprint arXiv:2312.06725}, 2023.

\bibitem[Kant et~al.(2024)Kant, Siarohin, Wu, Vasilkovsky, Qian, Ren, Guler,
  Ghanem, Tulyakov, and Gilitschenski]{kant2024spad}
Yash Kant, Aliaksandr Siarohin, Ziyi Wu, Michael Vasilkovsky, Guocheng Qian,
  Jian Ren, Riza~Alp Guler, Bernard Ghanem, Sergey Tulyakov, and Igor
  Gilitschenski.
\newblock Spad: Spatially aware multi-view diffusers.
\newblock In \emph{Proceedings of the IEEE/CVF Conference on Computer Vision
  and Pattern Recognition}, pp.\  10026--10038, 2024.

\bibitem[Karras et~al.(2022)Karras, Aittala, Aila, and Laine]{Karras2022edm}
Tero Karras, Miika Aittala, Timo Aila, and Samuli Laine.
\newblock Elucidating the design space of diffusion-based generative models.
\newblock In \emph{Proc. NeurIPS}, 2022.

\bibitem[Karras et~al.(2024)Karras, Aittala, Lehtinen, Hellsten, Aila, and
  Laine]{Karras2024edm2}
Tero Karras, Miika Aittala, Jaakko Lehtinen, Janne Hellsten, Timo Aila, and
  Samuli Laine.
\newblock Analyzing and improving the training dynamics of diffusion models.
\newblock In \emph{Proc. CVPR}, 2024.

\bibitem[Kawar et~al.(2023)Kawar, Zada, Lang, Tov, Chang, Dekel, Mosseri, and
  Irani]{kawar2023imagic}
Bahjat Kawar, Shiran Zada, Oran Lang, Omer Tov, Huiwen Chang, Tali Dekel, Inbar
  Mosseri, and Michal Irani.
\newblock Imagic: Text-based real image editing with diffusion models.
\newblock In \emph{Proceedings of the IEEE/CVF Conference on Computer Vision
  and Pattern Recognition}, pp.\  6007--6017, 2023.

\bibitem[Kingma(2013{\natexlab{a}})]{kingma2013auto}
Diederik~P Kingma.
\newblock Auto-encoding variational bayes.
\newblock \emph{arXiv preprint arXiv:1312.6114}, 2013{\natexlab{a}}.

\bibitem[Kingma(2013{\natexlab{b}})]{kingma2013vae}
Diederik~P Kingma.
\newblock Auto-encoding variational bayes.
\newblock \emph{arXiv preprint arXiv:1312.6114}, 2013{\natexlab{b}}.

\bibitem[Kong et~al.(2024)Kong, Liu, Lyu, Taher, Qi, and
  Davison]{kong2024eschernet}
Xin Kong, Shikun Liu, Xiaoyang Lyu, Marwan Taher, Xiaojuan Qi, and Andrew~J
  Davison.
\newblock Eschernet: A generative model for scalable view synthesis.
\newblock \emph{arXiv preprint arXiv:2402.03908}, 2024.

\bibitem[Li et~al.(2024{\natexlab{a}})Li, Ku, Yen, Liu, Liu, Chen, Kuo, and
  Sun]{li2024genrc}
Ming-Feng Li, Yueh-Feng Ku, Hong-Xuan Yen, Chi Liu, Yu-Lun Liu, Albert Chen,
  Cheng-Hao Kuo, and Min Sun.
\newblock Genrc: Generative 3d room completion from sparse image collections.
\newblock 2024{\natexlab{a}}.

\bibitem[Li et~al.(2024{\natexlab{b}})Li, Liu, Long, Zhang, Lin, Li, Qi, Zhang,
  Luo, Tan, et~al.]{li2024era3d}
Peng Li, Yuan Liu, Xiaoxiao Long, Feihu Zhang, Cheng Lin, Mengfei Li, Xingqun
  Qi, Shanghang Zhang, Wenhan Luo, Ping Tan, et~al.
\newblock Era3d: High-resolution multiview diffusion using efficient row-wise
  attention.
\newblock \emph{arXiv preprint arXiv:2405.11616}, 2024{\natexlab{b}}.

\bibitem[Li et~al.(2024{\natexlab{c}})Li, Pan, Yang, Xu, Zhou, Zhang, Li,
  Kadambi, Wang, and Fan]{Li20244K4DGenP4}
Renjie Li, Panwang Pan, Bangbang Yang, Dejia Xu, Shijie Zhou, Xuanyang Zhang,
  Zeming Li, Achuta Kadambi, Zhangyang Wang, and Zhiwen Fan.
\newblock 4k4dgen: Panoramic 4d generation at 4k resolution.
\newblock \emph{ArXiv}, abs/2406.13527, 2024{\natexlab{c}}.
\newblock URL \url{https://api.semanticscholar.org/CorpusID:270619480}.

\bibitem[Li et~al.(2023)Li, Zhang, and Ye]{li2023drivingdiffusion}
Xiaofan Li, Yifu Zhang, and Xiaoqing Ye.
\newblock Drivingdiffusion: Layout-guided multi-view driving scene video
  generation with latent diffusion model.
\newblock \emph{arXiv preprint arXiv:2310.07771}, 2023.

\bibitem[Liu et~al.(2024{\natexlab{a}})Liu, Li, Chen, Li, Xu, and
  Plummer]{liu2024panofree}
Aoming Liu, Zhong Li, Zhang Chen, Nannan Li, Yi~Xu, and Bryan~A Plummer.
\newblock Panofree: Tuning-free holistic multi-view image generation with
  cross-view self-guidance.
\newblock \emph{arXiv preprint arXiv:2408.02157}, 2024{\natexlab{a}}.

\bibitem[Liu et~al.(2024{\natexlab{b}})Liu, Wang, Liu, Bao, Han, and
  Yu]{liu2024mvpbev}
Buyu Liu, Kai Wang, Yansong Liu, Jun Bao, Tingting Han, and Jun Yu.
\newblock Mvpbev: Multi-view perspective image generation from bev with
  test-time controllability and generalizability.
\newblock \emph{arXiv preprint arXiv:2407.19468}, 2024{\natexlab{b}}.

\bibitem[Liu et~al.(2023{\natexlab{a}})Liu, Wu, Van~Hoorick, Tokmakov,
  Zakharov, and Vondrick]{liu2023zero}
Ruoshi Liu, Rundi Wu, Basile Van~Hoorick, Pavel Tokmakov, Sergey Zakharov, and
  Carl Vondrick.
\newblock Zero-1-to-3: Zero-shot one image to 3d object.
\newblock In \emph{Proceedings of the IEEE/CVF international conference on
  computer vision}, pp.\  9298--9309, 2023{\natexlab{a}}.

\bibitem[Liu et~al.(2023{\natexlab{b}})Liu, Lin, Zeng, Long, Liu, Komura, and
  Wang]{liu2023syncdreamer}
Yuan Liu, Cheng Lin, Zijiao Zeng, Xiaoxiao Long, Lingjie Liu, Taku Komura, and
  Wenping Wang.
\newblock Syncdreamer: Generating multiview-consistent images from a
  single-view image.
\newblock \emph{arXiv preprint arXiv:2309.03453}, 2023{\natexlab{b}}.

\bibitem[Long et~al.(2024)Long, Guo, Lin, Liu, Dou, Liu, Ma, Zhang, Habermann,
  Theobalt, et~al.]{long2024wonder3d}
Xiaoxiao Long, Yuan-Chen Guo, Cheng Lin, Yuan Liu, Zhiyang Dou, Lingjie Liu,
  Yuexin Ma, Song-Hai Zhang, Marc Habermann, Christian Theobalt, et~al.
\newblock Wonder3d: Single image to 3d using cross-domain diffusion.
\newblock In \emph{Proceedings of the IEEE/CVF Conference on Computer Vision
  and Pattern Recognition}, pp.\  9970--9980, 2024.

\bibitem[Lugmayr et~al.(2022)Lugmayr, Danelljan, Romero, Yu, Timofte, and
  Van~Gool]{lugmayr2022repaint}
Andreas Lugmayr, Martin Danelljan, Andres Romero, Fisher Yu, Radu Timofte, and
  Luc Van~Gool.
\newblock Repaint: Inpainting using denoising diffusion probabilistic models.
\newblock In \emph{Proceedings of the IEEE/CVF conference on computer vision
  and pattern recognition}, pp.\  11461--11471, 2022.

\bibitem[Lumalabs(2024)]{luma}
Lumalabs.
\newblock Dream machine., 2024.
\newblock URL \url{https://lumalabs.ai/dream-machine}.

\bibitem[Molad et~al.(2023)Molad, Horwitz, Valevski, Acha, Matias, Pritch,
  Leviathan, and Hoshen]{molad2023dreamix}
Eyal Molad, Eliahu Horwitz, Dani Valevski, Alex~Rav Acha, Yossi Matias, Yael
  Pritch, Yaniv Leviathan, and Yedid Hoshen.
\newblock Dreamix: Video diffusion models are general video editors.
\newblock \emph{arXiv preprint arXiv:2302.01329}, 2023.

\bibitem[M{\"u}ller et~al.(2024)M{\"u}ller, Schwarz, R{\"o}ssle, Porzi,
  Bul{\`o}, Nie{\ss}ner, and Kontschieder]{muller2024multidiff}
Norman M{\"u}ller, Katja Schwarz, Barbara R{\"o}ssle, Lorenzo Porzi,
  Samuel~Rota Bul{\`o}, Matthias Nie{\ss}ner, and Peter Kontschieder.
\newblock Multidiff: Consistent novel view synthesis from a single image.
\newblock In \emph{Proceedings of the IEEE/CVF Conference on Computer Vision
  and Pattern Recognition}, pp.\  10258--10268, 2024.

\bibitem[Pernias et~al.(2023)Pernias, Rampas, Richter, Pal, and
  Aubreville]{pernias2023wurstchen}
Pablo Pernias, Dominic Rampas, Mats~L Richter, Christopher~J Pal, and Marc
  Aubreville.
\newblock W{\"u}rstchen: An efficient architecture for large-scale
  text-to-image diffusion models.
\newblock \emph{arXiv preprint arXiv:2306.00637}, 2023.

\bibitem[Poole et~al.(2022)Poole, Jain, Barron, and
  Mildenhall]{poole2022dreamfusion}
Ben Poole, Ajay Jain, Jonathan~T. Barron, and Ben Mildenhall.
\newblock Dreamfusion: Text-to-3d using 2d diffusion.
\newblock \emph{arXiv}, 2022.

\bibitem[Rombach et~al.(2022)Rombach, Blattmann, Lorenz, Esser, and
  Ommer]{rombach2022high}
Robin Rombach, Andreas Blattmann, Dominik Lorenz, Patrick Esser, and Bj{\"o}rn
  Ommer.
\newblock High-resolution image synthesis with latent diffusion models.
\newblock In \emph{Proceedings of the IEEE/CVF conference on computer vision
  and pattern recognition}, pp.\  10684--10695, 2022.

\bibitem[Ruiz et~al.(2023)Ruiz, Li, Jampani, Pritch, Rubinstein, and
  Aberman]{ruiz2023dreambooth}
Nataniel Ruiz, Yuanzhen Li, Varun Jampani, Yael Pritch, Michael Rubinstein, and
  Kfir Aberman.
\newblock Dreambooth: Fine tuning text-to-image diffusion models for
  subject-driven generation.
\newblock In \emph{Proceedings of the IEEE/CVF conference on computer vision
  and pattern recognition}, pp.\  22500--22510, 2023.

\bibitem[Runway(2024)]{runway}
Runway.
\newblock Tools for human imagination., 2024.
\newblock URL \url{https://runwayml.com/product}.

\bibitem[Salimans \& Ho(2022)Salimans and
  Ho]{salimans2022progressivedistillation}
Tim Salimans and Jonathan Ho.
\newblock Progressive distillation for fast sampling of diffusion models, 2022.
\newblock URL \url{https://arxiv.org/abs/2202.00512}.

\bibitem[Shi et~al.(2023{\natexlab{a}})Shi, Chen, Zhang, Liu, Xu, Wei, Chen,
  Zeng, and Su]{shi2023zero123++}
Ruoxi Shi, Hansheng Chen, Zhuoyang Zhang, Minghua Liu, Chao Xu, Xinyue Wei,
  Linghao Chen, Chong Zeng, and Hao Su.
\newblock Zero123++: a single image to consistent multi-view diffusion base
  model.
\newblock \emph{arXiv preprint arXiv:2310.15110}, 2023{\natexlab{a}}.

\bibitem[Shi et~al.(2023{\natexlab{b}})Shi, Wang, Ye, Mai, Li, and
  Yang]{shi2023MVDream}
Yichun Shi, Peng Wang, Jianglong Ye, Long Mai, Kejie Li, and Xiao Yang.
\newblock Mvdream: Multi-view diffusion for 3d generation.
\newblock \emph{arXiv:2308.16512}, 2023{\natexlab{b}}.

\bibitem[Tang et~al.(2023)Tang, Zhang, Chen, Wang, and
  Furukawa]{Tang2023mvdiffusion}
Shitao Tang, Fuyang Zhang, Jiacheng Chen, Peng Wang, and Yasutaka Furukawa.
\newblock Mvdiffusion: Enabling holistic multi-view image generation with
  correspondence-aware diffusion.
\newblock \emph{arXiv}, 2023.

\bibitem[Tang et~al.(2024)Tang, Chen, Wang, Tang, Zhang, Fan, Chandra,
  Furukawa, and Ranjan]{tang2024mvdiffusion++}
Shitao Tang, Jiacheng Chen, Dilin Wang, Chengzhou Tang, Fuyang Zhang, Yuchen
  Fan, Vikas Chandra, Yasutaka Furukawa, and Rakesh Ranjan.
\newblock Mvdiffusion++: A dense high-resolution multi-view diffusion model for
  single or sparse-view 3d object reconstruction.
\newblock \emph{arXiv preprint arXiv:2402.12712}, 2024.

\bibitem[Unterthiner et~al.(2018)Unterthiner, van Steenkiste, Kurach, Marinier,
  Michalski, and Gelly]{Unterthiner2018ARXIV}
Thomas Unterthiner, Sjoerd van Steenkiste, Karol Kurach, Rapha{\"{e}}l
  Marinier, Marcin Michalski, and Sylvain Gelly.
\newblock Towards accurate generative models of video: {A} new metric {\&}
  challenges.
\newblock \emph{CoRR}, abs/1812.01717, 2018.

\bibitem[Valevski et~al.(2024)Valevski, Leviathan, Arar, and
  Fruchter]{valevski2024gamengen}
Dani Valevski, Yaniv Leviathan, Moab Arar, and Shlomi Fruchter.
\newblock Diffusion models are real-time game engines, 2024.
\newblock URL \url{https://arxiv.org/abs/2408.14837}.

\bibitem[Van~Hoorick et~al.(2024)Van~Hoorick, Wu, Ozguroglu, Sargent, Liu,
  Tokmakov, Dave, Zheng, and Vondrick]{van2024generative}
Basile Van~Hoorick, Rundi Wu, Ege Ozguroglu, Kyle Sargent, Ruoshi Liu, Pavel
  Tokmakov, Achal Dave, Changxi Zheng, and Carl Vondrick.
\newblock Generative camera dolly: Extreme monocular dynamic novel view
  synthesis.
\newblock \emph{arXiv preprint arXiv:2405.14868}, 2024.

\bibitem[Voleti et~al.(2024)Voleti, Yao, Boss, Letts, Pankratz, Tochilkin,
  Laforte, Rombach, and Jampani]{voleti2024sv3d}
Vikram Voleti, Chun-Han Yao, Mark Boss, Adam Letts, David Pankratz, Dmitry
  Tochilkin, Christian Laforte, Robin Rombach, and Varun Jampani.
\newblock Sv3d: Novel multi-view synthesis and 3d generation from a single
  image using latent video diffusion.
\newblock \emph{arXiv preprint arXiv:2403.12008}, 2024.

\bibitem[Wang et~al.(2023)Wang, Chen, Ling, Xie, and Song]{wang2023panodiff}
Jionghao Wang, Ziyu Chen, Jun Ling, Rong Xie, and Li~Song.
\newblock 360-degree panorama generation from few unregistered nfov images.
\newblock In \emph{Proceedings of the 31st ACM International Conference on
  Multimedia}, pp.\  6811--6821, 2023.

\bibitem[Wang \& Shi(2023)Wang and Shi]{wang2023imagedream}
Peng Wang and Yichun Shi.
\newblock Imagedream: Image-prompt multi-view diffusion for 3d generation.
\newblock \emph{arXiv preprint arXiv:2312.02201}, 2023.

\bibitem[Wang et~al.(2024{\natexlab{a}})Wang, Li, Mou, Cheng, and
  Zhang]{wang2024360dvd}
Qian Wang, Weiqi Li, Chong Mou, Xinhua Cheng, and Jian Zhang.
\newblock 360dvd: Controllable panorama video generation with 360-degree video
  diffusion model.
\newblock In \emph{Proceedings of the IEEE/CVF Conference on Computer Vision
  and Pattern Recognition}, pp.\  6913--6923, 2024{\natexlab{a}}.

\bibitem[Wang et~al.(2024{\natexlab{b}})Wang, Lv, Yu, Hong, Qi, Wang, Ji, Yang,
  Zhao, Song, Xu, Xu, Li, Dong, Ding, and Tang]{cogvlm}
Weihan Wang, Qingsong Lv, Wenmeng Yu, Wenyi Hong, Ji~Qi, Yan Wang, Junhui Ji,
  Zhuoyi Yang, Lei Zhao, Xixuan Song, Jiazheng Xu, Bin Xu, Juanzi Li, Yuxiao
  Dong, Ming Ding, and Jie Tang.
\newblock Cogvlm: Visual expert for pretrained language models,
  2024{\natexlab{b}}.
\newblock URL \url{https://arxiv.org/abs/2311.03079}.

\bibitem[Watson et~al.(2024)Watson, Saxena, Li, Tagliasacchi, and
  Fleet]{watson2024controlling}
Daniel Watson, Saurabh Saxena, Lala Li, Andrea Tagliasacchi, and David~J Fleet.
\newblock Controlling space and time with diffusion models.
\newblock \emph{arXiv preprint arXiv:2407.07860}, 2024.

\bibitem[Wen et~al.(2024)Wen, Zhao, Liu, Jia, Wang, Luo, Zhang, Wang, Sun, and
  Zhang]{wen2024panacea}
Yuqing Wen, Yucheng Zhao, Yingfei Liu, Fan Jia, Yanhui Wang, Chong Luo, Chi
  Zhang, Tiancai Wang, Xiaoyan Sun, and Xiangyu Zhang.
\newblock Panacea: Panoramic and controllable video generation for autonomous
  driving.
\newblock In \emph{Proceedings of the IEEE/CVF Conference on Computer Vision
  and Pattern Recognition}, pp.\  6902--6912, 2024.

\bibitem[Wu et~al.(2023)Wu, Zheng, and Cham]{wu2023panodiffusion}
Tianhao Wu, Chuanxia Zheng, and Tat-Jen Cham.
\newblock Panodiffusion: 360-degree panorama outpainting via diffusion.
\newblock In \emph{The Twelfth International Conference on Learning
  Representations}, 2023.

\bibitem[Wu et~al.(2024)Wu, Guo, Tang, Huang, Wang, Chen, and
  Ding]{wu2024drivescape}
Wei Wu, Xi~Guo, Weixuan Tang, Tingxuan Huang, Chiyu Wang, Dongyue Chen, and
  Chenjing Ding.
\newblock Drivescape: Towards high-resolution controllable multi-view driving
  video generation.
\newblock \emph{arXiv preprint arXiv:2409.05463}, 2024.

\bibitem[Yang et~al.(2024)Yang, Tan, Zhang, Wu, Li, Wetzstein, Liu, and
  Lin]{yang2024layerpano3d}
Shuai Yang, Jing Tan, Mengchen Zhang, Tong Wu, Yixuan Li, Gordon Wetzstein,
  Ziwei Liu, and Dahua Lin.
\newblock Layerpano3d: Layered 3d panorama for hyper-immersive scene
  generation.
\newblock \emph{arXiv preprint arXiv:2408.13252}, 2024.

\bibitem[Ye et~al.(2023)Ye, Zhang, Liu, Han, and Yang]{ye2023ip-adapter}
Hu~Ye, Jun Zhang, Sibo Liu, Xiao Han, and Wei Yang.
\newblock Ip-adapter: Text compatible image prompt adapter for text-to-image
  diffusion models.
\newblock 2023.

\bibitem[Yuan et~al.(2024)Yuan, Tang, Li, Yuille, and
  Wang]{yuan2024camfreediff}
Xiaoding Yuan, Shitao Tang, Kejie Li, Alan Yuille, and Peng Wang.
\newblock Camfreediff: Camera-free image to panorama generation with diffusion
  model.
\newblock \emph{arXiv preprint arXiv:2407.07174}, 2024.

\bibitem[Zhang et~al.(2024)Zhang, Wu, Gambardella, Huang, Phung, Ouyang, and
  Cai]{zhang2024taming}
Cheng Zhang, Qianyi Wu, Camilo~Cruz Gambardella, Xiaoshui Huang, Dinh Phung,
  Wanli Ouyang, and Jianfei Cai.
\newblock Taming stable diffusion for text to 360 $\{$$\backslash$deg$\}$
  panorama image generation.
\newblock \emph{arXiv preprint arXiv:2404.07949}, 2024.

\bibitem[Zhang et~al.(2023)Zhang, Rao, and Agrawala]{zhang2023controlnet}
Lvmin Zhang, Anyi Rao, and Maneesh Agrawala.
\newblock Adding conditional control to text-to-image diffusion models, 2023.

\bibitem[Zhang et~al.(2018)Zhang, Isola, Efros, Shechtman, and Wang]{lpips}
Richard Zhang, Phillip Isola, Alexei~A Efros, Eli Shechtman, and Oliver Wang.
\newblock The unreasonable effectiveness of deep features as a perceptual
  metric.
\newblock In \emph{CVPR}, 2018.

\bibitem[Zhao et~al.(2024)Zhao, Wang, Zhu, Chen, Huang, Bao, and
  Wang]{zhao2024drive}
Guosheng Zhao, Xiaofeng Wang, Zheng Zhu, Xinze Chen, Guan Huang, Xiaoyi Bao,
  and Xingang Wang.
\newblock Drivedreamer-2: Llm-enhanced world models for diverse driving video
  generation.
\newblock \emph{arXiv preprint arXiv:2403.06845}, 2024.

\bibitem[Zheng et~al.(2024)Zheng, Peng, Yang, Shen, Li, Liu, Zhou, Li, and
  You]{opensora}
Zangwei Zheng, Xiangyu Peng, Tianji Yang, Chenhui Shen, Shenggui Li, Hongxin
  Liu, Yukun Zhou, Tianyi Li, and Yang You.
\newblock Open-sora: Democratizing efficient video production for all, March
  2024.
\newblock URL \url{https://github.com/hpcaitech/Open-Sora}.

\bibitem[Zhou et~al.(2024)Zhou, Cheng, Yu, Tian, and Yuan]{zhou2024holodreamer}
Haiyang Zhou, Xinhua Cheng, Wangbo Yu, Yonghong Tian, and Li~Yuan.
\newblock Holodreamer: Holistic 3d panoramic world generation from text
  descriptions.
\newblock \emph{arXiv preprint arXiv:2407.15187}, 2024.

\end{thebibliography}
